\documentstyle[12pt,epsfig]{article}

\newlength{\dinwidth}
\newlength{\dinmargin}
\def\lapproxeq{\lower .7ex\hbox{$\;\stackrel{\textstyle
<}{\sim}\;$}}
\def\gapproxeq{\lower .7ex\hbox{$\;\stackrel{\textstyle
>}{\sim}\;$}}

\def\be{\begin{equation}}
\def\ee{\end{equation}}
\def\bea{\begin{eqnarray}}
\def\eea{\end{eqnarray}}

\def\GeV{\rm GeV}

\begin{document}
\titlepage
\begin{flushright}
IPPP/02/49 \\
DCPT/02/98 \\
Cavendish-HEP-2002/10 \\
5 November 2002 \\

\end{flushright}

\vspace*{0.5cm}

\begin{center}
{\Large \bf Uncertainties of predictions from parton
distributions\\ \vspace{1.5ex} I:\ Experimental errors}

\vspace*{1cm}
\textsc{A.D. Martin$^a$, R.G. Roberts$^b$, W.J. Stirling$^a$
and R.S. Thorne$^{c,}$\footnote{Royal Society University Research Fellow.}} \\

\vspace*{0.5cm} $^a$ Institute for Particle Physics Phenomenology,
University of Durham, DH1 3LE, UK \\
$^b$ Rutherford Appleton Laboratory, Chilton, Didcot, Oxon, OX11 0QX, UK \\
$^c$ Cavendish Laboratory, University of Cambridge, \\ Madingley Road,
Cambridge, CB3 0HE, UK
\end{center}

\vspace*{0.5cm}

\begin{abstract}
We determine the uncertainties on observables arising from the
errors on the experimental data that are fitted in the global
MRST2001 parton analysis. By diagonalizing the error matrix we
produce sets of partons suitable for use within the framework of
linear propagation of errors, which is the most convenient method
for calculating the uncertainties. Despite the potential
limitations of this approach we find that it can be made to work
well in practice. This is confirmed by our alternative approach of
using the more rigorous Lagrange multiplier method to determine
the errors on physical quantities directly. As particular examples
we determine the uncertainties on the predictions of the
charged-current deep-inelastic structure functions, on the
cross-sections for $W$ production and for Higgs boson production
via gluon--gluon fusion at the Tevatron and the LHC, on the ratio
of $W^-$ to $W^+$ production at the LHC and on the moments of the
non-singlet quark distributions. We discuss the corresponding
uncertainties on the parton distributions in the relevant $x,Q^2$
domains. Finally, we briefly look at uncertainties related to the
fit procedure, stressing their importance and using $\sigma_W$,
$\sigma_H$ and extractions of $\alpha_S(M_Z^2)$ as examples. As a
by-product of this last point we present a slightly updated set of
parton distributions, MRST2002.
\end{abstract}

\newpage

\section{Introduction}

Recently, much attention has been focused on uncertainties
associated with the parton distributions that are determined in
the next-to-leading order (NLO) global analyses of a wide range of
deep inelastic and related scattering data. There are many sources
of uncertainty, but they can be divided into two classes: those
which are associated with the {\em experimental} errors on the
data that are fitted in the global analysis and those which are
due to what can loosely be called {\em theory} errors. In this
latter category we have uncertainties due to (i)~NNLO and
higher-order DGLAP contributions, (ii)~effects beyond the standard DGLAP
expansion, such
as extra $\ln x$ and $\ln (1-x)$ terms, higher twist and
saturation contributions, (iii)~the particular choice of the
parametric form of the starting distributions, (iv)~heavy target
corrections, (v)~model assumptions, such as $s=\bar{s}$. In order to
estimate some of these `theory' errors, we can also look at
the uncertainties arising from
different choices of the data cuts ($W_{\rm cut}$, $x_{\rm cut}$,
$Q^2_{\rm cut}$), defined such that data with values of $W$, $x$
or $Q^2$ below the cut are excluded from the global fit. This approach
indicates where the current theory is struggling to fit the data compared
to other regions.

Here we study the uncertainties due to the errors on the data, and
leave the discussion of the `theory' uncertainties to a second
paper. Other groups \cite{Botje}--\cite{ZEUSfit} have also
concentrated on the experimental errors and have obtained
estimates of the uncertainties on parton distributions within a
NLO QCD framework, using a variety of competing procedures. Of
course, the parton distributions are not, themselves, physical
quantities. However, using the standard approach of the linear
propagation of errors, these uncertainties of the parton
distributions can be translated into uncertainties on observables.
Therefore, we first follow the general approach in \cite{CTEQHes} and
\cite{CTEQ6}, the Hessian method, and diagonalize the error
matrix, parameterizing an increase in $\chi^2$ of the fit in terms
of a quadratic function of the variation of the parameters away
from their best fit values. This gives us a number of sets of
partons with variations from the minimum in orthogonal directions
which can be used in a simple manner to calculate the uncertainty
on any physical quantity. However, this approach depends for its
reliability on the assumption that the quadratic dependence on the
variation of the parton parameters is very good. We find that this
approximation, with some modifications of the precise framework, i.e.
the elimination of some parameters and rescaling of others,
can be made to work well. We make available 30 sets of partons -- 2 for each
of the 15 eigenvector directions in parton parameter space --
which can be used to calculate the uncertainties on any physical quantity.

Despite its convenience, the Hessian approach does suffer from
some problems if one looks at it in detail, and if one tries to
extrapolate results, in particular if we consider large increases
in $\chi^2$. It is also not, in principle, the most suitable
method when allowing $\alpha_S$ to vary as one of the free
parameters in the fit. Hence, in this paper we also investigate
the uncertainties on observables directly. In order to do this we
apply the Lagrange multiplier method~\cite{CTEQLag} to the
observables themselves, therefore avoiding some of the
approximations involved in the linear propagation of errors from
partons to the observables, and confirming that these
approximations do not usually cause serious problems. When using
this Lagrange multiplier approach, the resulting sets of parton
distributions, which correspond to the extreme values of each
observable, can to a certain extent be thought of as the maximum
allowed variation of the dominant contributing partons in the
relevant kinematic ($x,Q^2$) domain. We select observables which
are particularly relevant for experiments at present and future
colliders, and which illustrate the uncertainties on specific
partons in a variety of kinematic ($x,Q^2$) domains. In order to
determine the true uncertainty on quantities we also let
$\alpha_S(M_Z^2)$ vary along with the parameters describing the
parton distributions directly, which is easy to implement in this
approach. Some quantities are then far more sensitive to
$\alpha_S(M_Z^2)$ than others. Fortunately our global fit
\cite{MRST2001} produces a value of $\alpha_S(M_Z^2)$ which is
consistent with the world average, with the same type of error,
i.e., $\alpha_S(M_Z^2)=0.119\pm0.002$. Hence, it is completely
natural to simply let $\alpha_S(M_Z^2)$ vary as a free parameter
in the fit in the same way as all the other parameters when
determining uncertainties. However, we also perform an
investigation of the uncertainties with $\alpha_S(M_Z^2)$ fixed at
0.119 in order to study more directly which variations in the
parton distributions are responsible for extreme variations in
given physical quantities, and to compare with the results of the
Hessian approach.

The physical observables that we select as examples in this
introductory study are, first, the charged-current structure
functions $F_2^{CC}(e^\pm p)$ for deep inelastic scattering at
high $x$ and $Q^2$ at HERA. These observables almost directly
represent the $d$, $u$ valence quarks at high $x$ and $Q^2$, where
deep inelastic data do exist \cite{H1A}--\cite{ZEUSCC}, but have
errors of $25\%$ or more at present. The precision on these data
is expected to increase dramatically in the near future. Second,
we determine the uncertainties on the cross-sections $\sigma_W$
and $\sigma_H$, for $W$ boson production and for the production of
a Higgs boson of mass\footnote{There is nothing special about the
choice of 115~GeV. We may choose different values in order to
probe the gluon in different $x,Q^2$ domains.} $M_H=115$~GeV by
gluon fusion respectively, at Tevatron and LHC energies. The cross
section $\sigma_W$ is sensitive to the sea quarks (and also, at
the Tevatron energy, weakly sensitive to the valence quarks) in a
range of rapidity centered about $x\sim M_W/\sqrt{s}$, and for
$Q^2\sim M_W^2$. Similarly, $\sigma_H$ is sensitive to the gluon
distribution in the domain $x\sim M_H/\sqrt{s}$ and $Q^2\sim
M_H^2$.

As a third example we determine the uncertainty on the ratio of
$W^-$ to $W^+$ production at the LHC energy. This ratio is
expected to be extremely accurately measured in the LHC
experiments. Other relevant examples, which we study, are the
uncertainties of the moments of the non-singlet ($u$--$d$) quark
distributions. These are quantities for which lattice QCD
predictions are becoming available, see, for example,
Refs.~\cite{LHPC-SESAM,QCDSF}.

The same techniques can be easily and quickly applied to a wide
variety of other physical processes sensitive to different partons
and different domains. Besides giving a direct evaluation of the
uncertainties on the observables, we can, in principle, unfold
this information to map out the uncertainties on NLO partons over
the whole kinematic domain where perturbative QCD is applicable.

The plan of the paper is as follows.
In Section~2 we discuss the Hessian method, and outline our
extraction of different parton distribution sets using this approach.
In particular we highlight the problems encountered, and how they are dealt
with in order to obtain reliable results. We make the sets of
partons obtained publicly available.
In Section~3 we briefly
recall the elements of the Lagrange multiplier method. In the
following four sections we determine the uncertainties of the
observables that we have mentioned above. This will involve a
series of global fits in which the observables are constrained at
different values in the neighbourhood of their values obtained in
the optimum global fit. In each case we explore, and discuss, the
allowed variation of the dominantly contributing partons. Using this more
rigorous method we also confirm
the general appropriateness of the Hessian approach, but discuss where it can
start to break down.

Finally, in Section~8, we summarize and briefly investigate the
uncertainties associated with the initial assumptions made in
performing the global fit. In order to do this we compare the $W$
and $H$ boson predictions with those obtained using both a
slightly updated set of our own partons, MRST2002, and using the
CTEQ6 partons~\cite{CTEQ6}. (All the results in Sections~2--7 are
based on MRST2001 partons~\cite{MRST2001}.) We find that for the
comparison with CTEQ some of the variations in predictions are
surprisingly large. We also illustrate the same result for the
extractions of $\alpha_S(M_Z^2)$ by various different groups. This
implies that uncertainties involved with initial assumptions and
also with theoretical corrections can be more important than those
due to errors on the data.

\section{The Hessian method}

The basic procedure involved in this approach is discussed in
detail in \cite{CTEQHes}, but we briefly introduce the important
points here. In this method one assumes that the deviation in
$\chi^2$ for the global fit\footnote{The data that are fitted can
be found in Refs.~\cite{H1Krakow,H1A,H1B} and
\cite{ZEUS}--\cite{Wasymm}. We treat the errors as in
\cite{MRST2001}.} from the minimum value $\chi_0^2$ is quadratic
in the deviation of the parameters specifying the input parton
distributions, $a_i$, from their values at the minimum, $a_i^0$.
In this case we can write
\begin{equation}
\Delta\chi^2=\chi^2-\chi_0^2 = \sum_{i=1}^{n}\sum_{j=1}^n H_{ij}(a_i-a^0_i)
(a_j-a^0_j),
\label{eq:hessian}
\end{equation}
where $H_{ij}$ is an element of the Hessian matrix, and $n$ is the number
of free input parameters. In this case the standard
linear propagation of errors allows one to calculate the error on any quantity
$F$ using the formula
\begin{equation}
(\Delta F)^2 = \Delta\chi^2
\sum_{i=1}^{n}\sum_{j=1}^n \frac{\partial F}{\partial a_i}
C_{ij}(a)\frac{\partial F}{\partial a_i},
\label{eq:covariance}
\end{equation}
where $C_{ij}(a) =  (H^{-1})_{ij}$ is the covariance, or error
matrix of the parameters, and $\Delta \chi^2$ is the allowed variation in
$\chi^2$. Hence, in principle, once one has either the Hessian or covariance
matrix (and a suitable choice of $\Delta\chi^2$) one can calculate the error
on any quantity.

However, as demonstrated in \cite{CTEQHes}, it
more convenient and more numerically stable to diagonalize either the Hessian
or covariance matrix, and work in terms of the eigenvectors. Since the
Hessian and covariance matrices are symmetric they have a set of orthogonal
eigenvectors defined by
\begin{equation}
\sum_{j=1}^n
C_{ij}(a)v_{jk} = \lambda_k v_{ik}.
\label{eq:eigeq}
\end{equation}
Moreover, because variations in some directions in parameter space lead to
deterioration in the quality of the fit far more quickly than others,
the eigenvalues $\lambda_k$
span several orders of magnitude. Hence it is helpful to work in
terms of the rescaled eigenvectors $e_{ik} = \sqrt{\lambda_k}v_{ik}$.
Then the parameter displacements from the minimum may be expressed as
\begin{equation}
\Delta a_i \equiv (a_i-a_i^0)= \sum_{k=1}^n e_{ik}z_k,
\label{eq:component}
\end{equation}
or using the orthogonality of the eigenvectors
\begin{equation}
z_i = (\lambda_i)^{-1}\sum_{k=1}^n e_{ki}\Delta a_k,
\label{eq:componentinv}
\end{equation}
i.e., the $z_i$ are the appropriately normalized combinations of
the $\Delta a_k$ which define the orthogonal directions in the
space of deviation of parton parameters. In practice a $z_i$ is
often dominated by a single $\Delta a_k$.\footnote{CTEQ have even
implemented the diagonalization procedure in the fitting procedure
itself in order to improve numerical stability \cite{CTEQmul}. We
do not think this will have effects significant enough to outweigh
the inherent errors in the Hessian approach described below.}

The error determination becomes much simpler in terms of the $z_i$.
The increase in $\chi^2$ is
\begin{equation}
\Delta\chi^2= \sum_{i=1}^{n}z_i^2,
\label{eq:hessiandiag}
\end{equation}
i.e., the surface of constant $\chi^2$ is a hyper-sphere of given
radius in $z$-space. Similarly the error on the quantity $F$ is
now
\begin{equation}
\Delta F = \sqrt{\Delta\chi^2}\,
\biggl[\, \sum_{i=1}^{n}\biggl(\frac{\partial F}{\partial z_i}
\biggr)^2\,\biggr]^{1/2}.
\label{eq:covariancediag}
\end{equation}
Thus it is convenient to introduce parton sets $S^{\pm}_k$ for
each eigenvector direction, i.e., from Eq.~(\ref{eq:component}) we
define
\begin{equation}
\Delta a_i(S_k^{\pm}) = \pm  t e_{ik},
\label{eq:setk}
\end{equation}
where  the tolerance $t$ is defined by
$t=\sqrt{\Delta \chi^2}$ and $\Delta \chi^2$ is the allowed
deterioration in fit quality for the error determination.
Then, assuming the quadratic behaviour of $F$ about the minimum,
(\ref{eq:covariancediag}) becomes the simple expression
\begin{equation}
(\Delta F) = \frac{1}{2}\left[\,\sum_{k=1}^{n}
\left(F(S_k^+)-F(S_k^-)\right)^2\,\right]^{\frac{1}{2}}.
\label{eq:covdiag}
\end{equation}

If everything were ideal this framework would provide us with a simple
and efficient method for calculating the uncertainty due to experimental
errors on any quantities, where we would use the standard choice of
$\Delta \chi^2=1$. However, the real situation is not so simple,
and there are two major complications we must overcome in order to obtain
reliable results.

Although, in principle, the $1\sigma$ uncertainty in any
cross-section should be given by $\Delta\chi^2=1$, the complicated
nature of the global fitting procedure, where a large number of
independent data sets are used, results in this being an
unrealistically small uncertainty~\cite{THJPG}. This is
undoubtedly due to some failure of the theoretical approximation
to work absolutely properly over the full range of data, which
introduces the type of theoretical errors outlined in the
Introduction, and also due to some sources of experimental error
not being precisely quantified. Both problems are in practice
extremely difficult to surmount. We shall implicitly ignore the
potential theoretical error in this paper, but account for the
lack of ideal behaviour in the framework by determining the
uncertainties using a larger $\Delta \chi^2$. We estimate
$\Delta\chi^2=50$ to be a conservative uncertainty (perhaps of the
order of a $90\%$ confidence level or a little less than $2\sigma$) due to the
observation that an increase of 50 in the global $\chi^2$, which
has a value $\chi^2= 2328$ for $2097$ data points, usually
signifies that the fit to one or more data sets is becoming
unacceptably poor. We find that an increase $\Delta\chi^2$ of 100
normally means that some data sets are very badly described by the
theory. Though this estimation does not rely on any real
mathematical foundation we do not think it is any less valid than
the approaches used in e.g. \cite{CTEQ6} or \cite{Botje,ZEUSfit},
both of which ultimately appeal to some value judgment rather than
using all available information in a statistically rigorous
manner, and ultimately give similar results. The approaches
\cite{Giele, Alekhin, H1Krakow} do use $\Delta\chi^2=1$ but either
rely on much smaller and more internally compatible data sets, or
in some cases have rather small errors.

The second complication is the breakdown of the simple quadratic
behaviour in terms of variations of the parameters, i.e., the fact
that Eq.~(\ref{eq:hessian}) may receive significant corrections
and the simple linear propagation of errors is therefore not
accurate. Of course, we expect some deviations from this simple
form for very large $\Delta \chi^2$, but unfortunately very
significant deviations can occur for relatively small $\Delta
\chi^2$, as outlined below. Due to the very large amount of data
in our global fit, we have a lot of parameters in order to allow
sufficient flexibility in the form of the parton distributions.
Each of the valence quarks and the total sea quark contribution
are parameterized in the form
\begin{equation}
xq(x,Q_0^2)=A(1-x)^{\eta}(1+\epsilon x^{0.5}+\gamma x)x^{\delta},
\label{eq:quarks}
\end{equation}
where for the valence quarks the normalization $A$ is set by the number of
valence quarks of each type.
Because we find it necessary to have a negative input gluon at low $x$
the gluon parameterization has been extended to
\begin{equation}
xg(x,Q_0^2)=A_g(1-x)^{\eta_g}(1+\epsilon_gx^{0.5}+\gamma_g x)x^{\delta_g}
-A_-(1-x)^{\eta_-}x^{-\delta_-},
\label{eq:gluon}
\end{equation}
where $A_g$ is determined by the momentum sum rule, and $\eta_-$
can be set to some fixed large value, e.g. 10 or 20, so that the
second term only influences large $x$. The combination $\Delta q =
\bar u - \bar d$ has a slightly different parameterization, i.e.,
\begin{equation}
x\Delta q(x,Q_0^2)=A(1-x)^{\eta}(1+\gamma x+\delta x^2)x^{\delta}.
\label{eq:quarkdif}
\end{equation}
Overall, this gives 24 free parameters. In our standard fits we
allow all these parameters to vary. However, when investigating in
detail the small departures from the global minimum we notice that
a certain amount of redundancy in parameters leads to potentially
disastrous departures from the behaviour in
Eq.~(\ref{eq:hessian}). For example, in the negative term in the
gluon parameterization very small changes in the value of
$\delta_-$ can be compensated almost exactly by a change in $A_-$
and (to a lesser extent) in the other gluon parameters over the
range of $x$ probed, and therefore changes in $\delta_-$ lead to
very small changes in $\chi^2$. However, at some point the
compensation starts to fail significantly and the $\chi^2$
increases dramatically. Hence, this certain degree of redundancy
between $\delta_-$ and $A_-$ leads to a severe breaking of the
quadratic behaviour in $\Delta \chi^2$. Essentially the redundancy
between the parameters leads to a very flat direction in the
eigenvalue space (a very large/small eigenvalue of the
covariance/Hessian matrix) which means that cubic, quartic {\it
etc}. terms dominate. During the process of diagonalization this
bad behaviour feeds through into the whole set of eigenvectors to
a certain extent.

Therefore, in order that the Hessian method work at all well we
have to eliminate the largest eigenvalues of the covariance
matrix, i.e., remove the redundancy from the input parameters. In
order to do this we simply fix some of the parameters at their
best fit values so that the Hessian matrix only depends on a
subset of parameters that are sufficiently independent that the
quadratic approximation is reasonable. In fact we finish up with
15 free parameters in total -- 3 for each of the 5 different types
of input parton. In particular, fixing the other parameters at the
best fit values we find that $\eta_g$, $\delta_g$ and $\delta_-$
are sufficient for the gluon -- one for high $x$, one for medium
$x$ and one for low $x$. However, we emphasize that we cannot
simply set the other parameters to zero. For example $A_-$ must be
of a size as to allow a sufficiently negative input gluon at low
$x$ with a sensible value of $\delta_-$, but we cannot allow it to
vary simultaneously with $\delta_-$. We could possibly allow one
or two more parameters to be free, but judge that the
deterioration in the quality of the quadratic approximation does
not outweigh the improvements due to increased flexibility in the
parton variations. We note that this problem seems to be a feature
of the full global fits obtained by CTEQ and MRST, and that the
other fitting groups have not yet needed to introduce enough
parameters to notice such redundancy. It has clearly been noticed
by CTEQ though, since in \cite{CTEQHes} they only have 16 free
parameters out of a possible 22, and in \cite{CTEQ6}, where they
use a significantly altered type of parameterization, they have
only 20 free parameters out of a possible 26.

Hence, we produce 30 sets of parton distributions labeled by
$S^{\pm}_k$ to go along with the central best fit; that is 15
``+'' sets corresponding to each eigenvector direction, and 15
``--'' sets\footnote{In order to produce the errors on the parton
distributions a higher numerical accuracy was required than that
used when we previously found just the ``best fit''. This results
in the partons from the central fit being very slightly different
to the standard MRST2001 partons, and we label them by MRST2001C.
In fact some of the input parameters are quite different to those
in the MRST2001 default, but the partons themselves differ by
fractions of a percent. This is an example of the redundancy in
some input parameters noted above. The 31 parton sets ($S_k^\pm$,
MRST2001C) are available at
http://durpdg.dur.ac.uk/hepdata/mrs$\,$.}. Even though we have
limited the number of free parameters in the calculation of the
Hessian matrix, we note that we still have significant departure
from the ideal quadratic behaviour. For the 10 or so lowest
eigenvalues of the covariance matrix the quadratic approximation
is very good -- the distance needed to go along one of the $z_i$
to produce $\Delta \chi^2=50$ being the expected $\sqrt{50} =
7.07$ to good accuracy in both ``+'' and ``--'' directions.
However, for 4 or 5 of the largest eigenvalues of the covariance
matrix, corresponding mainly to the large-$x$ $d$ quark, large-$x$
gluon and $\bar u - \bar d$ distributions, the absolute scaling
and symmetry break down somewhat. In the very worst case of the
largest eigenvalue, the scale factors to produce $\Delta \chi^2
=50$ are 9.5 and 4.5 in the two opposite directions. In order to
produce the sets corresponding to $\Delta \chi^2 =50$ we have to
multiply the parton deviations required for $\Delta \chi^2 =1$ by
these scale factors rather than the expected 7.07. (In fact we do
this for all 30 sets, but in most cases the scale factor is in the
range 6.5--7.5.) Hence, as in \cite{CTEQHes,CTEQ6}, this
necessitates the supply of both ``+'' and ``--'' sets, whereas in
the quadratic approximation one could easily be obtained from the
other. Indeed from Fig.~9 of \cite{CTEQHes} it is clear that CTEQ
encounter a breakdown of the quadratic behaviour of much the same
type that we do.

Using the 30 parton sets $S_k^\pm$ corresponding to the 15
eigenvector directions for variations of the partons about the
minimum $\chi^2$, one can use Eq.~(\ref{eq:covariancediag}) to
calculate the error for any quantity, assuming an allowed $\Delta
\chi^2 =50$. In fact it has been proposed \cite{sullnad} that one may 
also account for 
some of the asymmetry due to departures from quadratic behaviour by 
replacing Eq.~(\ref{eq:covdiag}) by the slightly more sophisticated form 
\bea
(\Delta F)_+ &= \left[\,\sum_{k=1}^{n}
\left( \max(F(S_k^+)-F(S_k^0),F(S_k^-)-F(S_k^0),0)\right)^2\,
\right]^{\frac{1}{2}} \nonumber \\
(\Delta F)_- &= \left[\,\sum_{k=1}^{n}
\left( \max(F(S_k^0)-F(S_k^+),F(S_k^0)-F(S_k^-),0)\right)^2\,
\right]^{\frac{1}{2}},
\label{eq:covdiagasymm}
\eea 
where $S_k^0$ represents the best fit set of partons. In \cite{sullnad} 
and \cite{sull} examples are discussed where the use of 
Eq.~(\ref{eq:covdiagasymm}) instead of
Eq.~(\ref{eq:covdiag}) leads to not only an asymmetric error, but also a larger
uncertainty overall. We find only fairly minor effects, with no real evidence 
that Eq.~(\ref{eq:covdiagasymm}) leads to markedly more reliable results than
Eq.~(\ref{eq:covdiag}), so we use the simpler 
Eq.~(\ref{eq:covdiag}) in this paper. 

As an example of the use of the Hessian method we show in
Figs.~1--4 the uncertainty on some of the parton distributions at
various values of $Q^2$, namely the $u_V$ distribution, the $d_V$
distribution and the gluon distribution respectively. As one sees,
the $u_V$ distribution is very well determined, and the
uncertainty shrinks slightly with increasing $Q^2$. The lowest
uncertainty is in the region of $x=0.2$ where there are very
accurate data which mainly constrain the valence quarks. At lower
$x$ the direct constraint is on the sum of valence and sea quarks.
The $d_V$ distribution is also well determined in general, but is
rather more uncertain as we go to the highest $x$ values. The
gluon distribution is known less well, but at the highest $Q^2$
has an uncertainty of as little as $5\%$ for $x \sim 0.05$ where
it is constrained by both $dF_2(x,Q^2)/d\ln Q^2$ of the HERA data
and the lowest-$E_T$ Tevatron jet data. It becomes very uncertain
for $x\geq 0.4$ where only the relatively imprecise highest-$E_T$
jet data provide any information. The fractional uncertainty at
very small $x$ decreases very rapidly as $Q^2$ increases because
much of the small-$x$ gluon at higher $Q^2$ is generated from that
at higher $x$ via evolution. We also show the gluon at $Q^2=2\
\GeV^2$ explicitly in Fig.~4. At this low scale the central gluon
is negative at $x=0.0001$, but we see that the gluon may be
positive within the uncertainty. This just about persists if we go
to as low as $x=10^{-5}$ at this $Q^2$, but at our input scale
$Q_0^2=1\ \GeV^2$ the gluon would be negative for $x$ less than
0.0005, outside the level of uncertainty chosen. Also shown on the
plots are the CTEQ6M partons. For the $d_V$ distribution the
agreement is excellent. For the $u_V$ distribution the agreement
at $x\geq 0.05$ is very good, but there is a discrepancy below
this value. However, in this range, the valence quarks become very
small indeed and the data only really constrain the total $u$
distribution which is completely dominated by the sea. This
apparent discrepancy is probably due to parameterization effects,
and is irrelevant in practice. However, in Fig.~3 we see that the
MRST and CTEQ gluons show a genuine and significant level of
incompatibility. We will comment on this more in Section~8.

One might worry that the fixing of some of the parameters, that
determine the input parton distributions, will cast some doubt on
the error obtained. However, we stressed that these are largely
redundant parameters, and we have checked that the errors obtained
(when using $\Delta \chi^2 =50$) are indeed compatible with the
errors obtained using more rigorous means, i.e., the Lagrange
multiplier method, in the following sections.\footnote{We have checked the 
effects of using Eq.~(\ref{eq:covdiagasymm}) instead of
Eq.~(\ref{eq:covdiag}) in these comparisons. In all cases the former only 
introduced a relatively small asymmetry in the uncertainty, with the 
average being very close indeed to the result using the latter. 
Also, the asymmetry
was of the same sign as that found using the Lagrangian approach only 
half the time, i.e. the use of Eq.~(\ref{eq:covdiagasymm}) did not reliably 
predict the direction of steeper increase of $\Delta \chi^2$, even 
when the asymmetry was quite large. We find this
surprising, and have no good explanation. However, it illustrates the 
semi-qualitative nature of the Hessian approach compared to the 
more rigorous Lagrange Multiplier method.} 
Nevertheless, it is a sign of the breakdown of the quadratic 
approximation. Of more
practical concern is the fact that this breakdown is also
exhibited in a non-trivial manner in some of the eigenvectors used
-- particularly those eigenvectors associated with the least known
partons, e.g. the high-$x$ down quark and gluon. The scaling has
been designed to give correct results if $\Delta \chi^2 =50$ is
used. However, one cannot simply extrapolate to different choices
of $\Delta \chi^2$. For example if $\Delta \chi^2 =25$ were deemed
a more suitable choice, in principle the error would just be that
using Eq.~(\ref{eq:covariancediag}) divided by $\sqrt{2}$, but the
breakdown of quadratic behaviour does not guarantee this,
especially for some directions in parameter space. Also, if one
wished to be very conservative in the estimation of an
uncertainty, simple extrapolation cannot reveal when $\Delta
\chi^2$ might start to increase rapidly. We will see examples of
this later.

Also we note that we have performed this analysis for a fixed
value of the coupling constant: $\alpha_S(M_Z^2)=0.119$. One can
in principle include this as another free parameter. Indeed we
then find that the behaviour obeys the quadratic approximation
quite well and that $\Delta \chi^2 =50$ gives an error of about
$\pm 0.003$, corresponding well to our error of $\pm0.002$
obtained in \cite{MRST2001} using $\Delta \chi^2 =20$.
We will discuss extractions of $\alpha_S(M_Z^2)$ again in Section~8.
However, for the Hessian approach there is a slight difference between
varying $\alpha_S(M_Z^2)$ and varying the parton parameters.
When $\alpha_S(M_Z^2)$ is fixed the maximum error on any quantity is
obtained from some linear combination of our different parton
sets, and in principle one could reproduce the particular parton
set which corresponds to this linear combination, which would be a
perfectly well-defined set itself. However, a linear combination
of $\alpha_S(Q^2)$ coming from contributions with different
$\alpha_S(M_Z^2)$ does not actually correspond to one particular
choice of $\alpha_S(M_Z^2)$ (each contribution has a branch point
at a different value of $Q^2$, so a linear combination will have
multiple branch points), so one cannot precisely define a
particular set of partons corresponding to a particular
$\alpha_S(M_Z^2)$ for the extreme.

Hence, although the 30 parton sets obtained using the Hessian
approach provide the most convenient framework for calculating the
uncertainties on a physical observable, for the reasons described
above we would also like to study an alternative approach,
partially just to check how well our adapted Hessian approach
really works. A more robust method, which also allows us to
directly investigate the partons, and $\alpha_S$, corresponding to
the extreme variations of a given physical quantity is the
Lagrange multiplier method. We study this in detail below.

\section{Lagrange multiplier method}

It is much more rigorous to investigate the allowed variation of a
specific observable by using the Lagrange multiplier method. This
was also one of the approaches used by the CTEQ
collaboration~\cite{CTEQLag}. In this, one performs a series of
global fits while constraining the values $\sigma_i$ of one, or
more, physical quantities in the neighbourhood of their values
$\sigma_i^0$ obtained in the unconstrained global fit.  To be
precise, we minimize the function \be \Psi(\lambda_i,a)\ =\
\chi_{\rm global}^2(a) + \sum_i\lambda_i\sigma_i(a)\label{eq:Psi}
\ee with respect to the usual set $a$ of parameters, which specify
the parton distributions and the coupling $\alpha_S(M_Z^2)$. This
global minimization is repeated for many fixed values of the
Lagrange multipliers $\lambda_i$. At the minima, with the lowest
$\Psi(\lambda_i,a)$, the observables have the values
$\sigma_i(\widehat{a})$ and the value of $\chi_{\rm
global}^2(\widehat{a})$ is the minimum for these particular values
of $\sigma_i$. These optimum parameter sets $\widehat{a}$ depend
on the fixed values of $\lambda_i$. Clearly, when $\lambda_i=0$,
we have $\Psi=\chi_{\rm global}^2=\chi_0^2$ and
$\sigma_i=\sigma_i^0$. In this way we are able to explore how the
global description of the data deteriorates as the
$\sigma_i(\widehat{a})$ move away from the unconstrained best fit
values $\sigma_i^0$. Thus by spanning a range of $\lambda_i$ we
obtain the $\chi_{\rm global}^2$ profile for a range of values of
$\sigma_i$ about the best fit values, $\sigma_i^0$. In this study
we take the best fit values corresponding to the MRST2001
partons~\cite{MRST2001}.

This procedure involves none of the approximations involved in the
Hessian approach. We can use the full set of parameters in the
fit, obtaining maximum flexibility in the partons without having
to worry about the large correlations or anticorrelations between
some parameters. We never make any assumption about quadratic
dependence on the parameters, and indeed, by using different
values of the Lagrange multipliers, we can map out precisely how
the quadratic approximation breaks down in the uncertainty for any
physical quantity. Also, one produces a particular set of partons
with a particular value of $\alpha_S(M_Z^2)$ at every point in the
space of cross-sections for the physical quantities mapped, so the
interpretation of the extremes is more obvious and natural. Hence,
in principle, this is a far superior method of obtaining
uncertainties to the Hessian approach. However, it suffers from
the large practical disadvantage that a series of global fits must
be done every time one considers a new quantity. As examples we
investigate a number of interesting physical cases below.

\section{The charged-current structure functions $F_2^{CC}(e^\pm p$)}

The $\Delta\chi^2$ contour plot for the variation of $F_2^{CC}(e^+
p)$ and $F_2^{CC}(e^- p)$ about their predicted values from the
unconstrained global fit is shown in Fig.~5, where we allow
$\alpha_S$ to be a free parameter (unstarred labels) or fix it at
the best fit value of $\alpha_S(M_Z^2)=0.119$ (starred labels). We
show the contours for $\Delta\chi^2=50$,~100,~etc. Overall, the
ellipses one would expect from the quadratic approximation for
$\Delta \chi^2$ in Section~2 are more or less what one sees, but
there is a certain asymmetry in that $\chi^2$ increases rather
more rapidly for an increase in both $F_2^{CC}(e^+p)$ and
$F_2^{CC}(e^- p)$ than for a corresponding decrease in both.

Thus, from Fig.~5, we see that the uncertainties of the
$F_2^{CC}(e^+ p)$ and $F_2^{CC}(e^- p)$ structure functions at
$x=0.5$ and $Q^2=10,000\ \GeV^2$ (due to the experimental errors
on the data in the global fit) are about
\raisebox{-1ex}{$\stackrel{\textstyle +15}{\footnotesize -12}$}\
\% and $\pm2\%$ respectively. In comparison, the values using the
Hessian approach are $\pm 10\%$ and $\pm 2\%$ respectively, in
good agreement, although slightly smaller for $F_2^{CC}(e^+ p)$.
At this value of $x$ the uncertainties in $F_2^{CC}(e^+ p)$ and
$F_2^{CC}(e^- p)$ have a particularly simple interpretation since
$F_2^{CC}(e^+ p)$ is almost exactly proportional to the valence
down quark distribution, $d_V$, and $F_2^{CC}(e^- p)$ is almost
exactly proportional to the $u_V$ distribution. This can clearly
be seen in Fig.~6, which shows the $u$ and $d$ distributions for
the extreme sets (T*,U*,V* and W*) corresponding to maximum and
minimum $F_2^{CC}(e^+ p)$ and $F_2^{CC}(e^- p)$ (for the case of
fixed $\alpha_S(M_Z^2)$). Rather obviously the $d$ distribution
maximises at large $x$ for the case of maximum $F_2^{CC}(e^+ p)$
and minimises for minimum $F_2^{CC}(e^+ p)$, with similar
behaviour for the $u$ distribution and $F_2^{CC}(e^- p)$. Note
however that in each case the extreme in the parton distribution
is not precisely at $x=0.5$, but at slightly higher $x$, where the
data are less constraining. There are also sum rules on the
partons which must be satisfied. It is also clear that there is a
strong inverse correlation between the $u$ and $d$ distributions.
This is because the data which constrain the relevant partons are
the structure function measurements $F_2 (l p)$, $F_2(l d)$ and
$F_{2(3)}(\nu(\bar\nu)  p)$ which are essentially proportional to
$4u+d$, $u+d$ and $u+d$ respectively, where $u\sim 4d$ at $x=0.5$.
This constrains $u$ far more than $d$ as we have seen, but means
that for maximum variation in the partons a change in $u$ must be
compensated by a much larger opposite change in $d$. The result
that the major axis of the ellipse for given change in $\Delta
\chi^2$ is approximately aligned along $8F_2^{CC}(e^+
p)-F_2^{CC}(e^- p)$ (i.e., $8d-u$) is therefore not at all
surprising. The rate of quickest increase in $\chi^2$ is then
along $8d+u$, where the changes in the partons add in such a way
as to maximise changes in the measured structure functions.

We see that allowing $\alpha_S(M_Z^2)$ to also vary allows the
error ellipses to grow slightly, mainly in width. Now the maximum
and minimum allowed values of $F_2^{CC}(e^- p)$ (or $u$)
correspond to $\alpha_S(M_Z^2)=0.117$ and $0.120$ and to parton
sets T and V respectively. Most of the constraining data are for
$Q^2 \ll 10,000\ \GeV^2$, and must be well fit, but smaller
$\alpha_S$ means slower evolution of the quarks and thus greater
values of $u$ and $F_2^{CC}(e^- p)$ at $Q^2=10,000\ \GeV^2$.
Opposite considerations lead to the maximum $F_2^{CC}(e^- p)$.
Since the extrema of $F_2^{CC}(e^+ p)$ and $d$ are more involved,
due to the negative correlation with the $u$ distribution, they
are less altered by allowing $\alpha_S$ to vary; see sets U and W.
We see that the axes of the ellipse are essentially unchanged when
$\alpha_S$ is left free. Thus Fig.~6 is much the same except that
the variations for parton sets T and V are a little greater than
for T$^*$ and V$^*$.

It is, of course, the fixed target data which constrain these
cross-sections and the high-$x$ quarks. It is very largely the
BCDMS $F_2(e d)$ measurements which are responsible for the upper
extremum in $F_2^{CC}(e^+ p)$. The best fit tends to overshoot
these data in the region of $x=0.5$, and a large increase in $d$
makes the fit to these measurements very poor. For the extrema in
$F_2^{CC}(e^- p)$ and $u$, the deterioration is more evenly spread
over pretty much all the fixed target data at $x\simeq0.5$ (with
the exception that the description of the BCDMS $F_2(d)$
measurements improves slightly), but the cumulative result is a
very poor fit. One of the worst instances of deterioration is for
the NMC $F_2(n)/F_2(p)$ ratio.

\section{$W$ and $H$ production at the LHC and Tevatron}

The $\Delta\chi^2$ contour plot for the variation of $\sigma_W$
and $\sigma_H$ about their predicted values at the LHC energy from
the unconstrained global fit is shown in Fig.~7, where again we
allow $\alpha_S$ to be a free parameter or fix it at
$\alpha_S(M_Z^2)=0.119$. Again we show the contours for
$\Delta\chi^2=50$,~100,~etc. This time the Hessian approach should
work well, although the ellipses start becoming a little
rectangular. Allowing $\alpha_S(M_Z^2)$ to vary, we see that the 
uncertainties of the $W$ and $H$
cross-sections at the LHC (due to the experimental errors on the
data in the global fit) are about
\raisebox{-1ex}{$\stackrel{\textstyle +2.5}{\footnotesize -2.0}$}\
\% and $\pm 3\%$ respectively, and are positively correlated.

Again this analysis also gives information on the uncertainties of
particular parton distributions. To be specific, the parton sets
which correspond to the points A,B,C,D, on the $\Delta\chi^2=50$
contour in Fig.~7, give the uncertainties in the parton
distributions that dominantly determine $\sigma_W$ and $\sigma_H$
in the kinematic domain $x\sim0.005$, $Q^2\sim10^4$~GeV$^2$
relevant to $W$ and $H$ production at the LHC. The extrema in
$\sigma_W$, represented by A and C, correspond to variations in
the sea quark distributions, while the extrema in $\sigma_H$,
represented by B and D, correspond to variations in the gluon
distribution {\em and} $\alpha_S(M_Z^2)$. The values of $\alpha_S$
for sets A and C are 0.119 and 0.118 respectively, both very close
to the default MRST2001 value, showing that $\sigma_W$, which
begins at zeroth order, is insensitive to $\alpha_S$. However, the
values of $\alpha_S$ for fits B and D are 0.120 and 0.117
respectively, reflecting the fact that
$\sigma_H\propto\alpha_S^2$. This is well illustrated by repeating
the entire analysis with $\alpha_S$ fixed at the default value
(0.119) obtained in the unconstrained global fit \cite{MRST2001}.
The $\Delta\chi^2=50$ and 100 contours for this additional
analysis are shown by the smaller shapes in Fig.~7. We can see
that the uncertainty on $\sigma_W$ is almost unchanged, while that
for $\sigma_H$ is reduced to about $\pm 2\%$.  The
corresponding values using the Hessian approach are $\pm 1.8\%$ and
$\pm 1.8\%$, in good agreement but slightly smaller in each case.

The up quark distribution for each `extrema' set with fixed
$\alpha_S(M_Z^2)$ is shown in Fig.~8(a) and the gluon distribution
in Fig.~8(b). We see that indeed the parton distributions do
reflect the extrema in the cross-sections in a fairly simple
manner. The quark densities at high $x$ show almost no variation
between fits since they are well constrained at high $x$ and
because the $W$ and $H$ production cross-sections are sensitive to
the partons at an $x$ range centered at a few $\times 10^{-3}$.
Indeed the maximum and minimum $W$ cross-sections correspond to
the maximum and minimum sea quarks for $x \leq 0.05$ at $Q^2 \sim
10,000\ \GeV^2$. The maximum and minimum Higgs cross-sections
correspond to the maximum and minimum gluon distributions in the
same sort of range, although the large $x$ gluon must now decrease
for increases in the small $x$ partons, and {\it vice versa}, in
order to maintain the momentum sum rule. The strong correlation
between the two cross-sections is due to the fact that at high
$Q^2$ the size of the quark distribution at small $x$ is mainly
determined by evolution, and the larger the small $x$ gluon the
stronger the quark evolution (and {\it vice versa}). When
$\alpha_S$ is left free the resulting partons at the extrema are
similar to the fixed $\alpha_S$ results. However, in this case,
their variation is a little larger at smaller $Q^2$, since the
slight changes in $\alpha_S$ lead to different rates of evolution.

For the case of fixed $\alpha_S$ the main contributions to $\Delta
\chi^2$ come from the HERA small-$x$ structure function data and,
because of the changes in the high $x$ gluon, also from the
Tevatron jet data. For the upper extrema in $\sigma_W$ and
$\sigma_H$ the slope $dF_2(x,Q^2)/d\ln Q^2$ tends to be too large
for $x \leq 0.001$, while for the lower extrema the slope is too
small. In both cases the fit to jet data deteriorates due to the
shape of the high-$x$ gluon becoming wrong. When $\alpha_S(M_Z^2)$
is allowed to vary the data which are particularly sensitive to
this also play a role, for example the BCDMS data are fitted less
well when $\alpha_S(M_Z^2)=0.120$ in fit B, and the NMC data are
described less well when $\alpha_S(M_Z^2)=0.117$ in fit~D.

\bigskip

The corresponding $\Delta\chi^2$ contour plot for the Tevatron is
shown in Fig.~9, where again we either allow $\alpha_S$ to be a
free parameter or fix it at $\alpha_S(M_Z^2)=0.119$. We see that
the uncertainty of the $W$ cross-section at the Tevatron (due to
the experimental errors on the data in the global fit) has
decreased to about $\pm1.5\%$ while that for the Higgs has
increased to about $\pm8$\% for varying $\alpha_S(M_Z^2)$, and
that the correlation has disappeared. For $\alpha_S(M_Z^2)$ fixed
at 0.119 $\sigma_W$ is again largely unaffected, but the uncertainty of
$\sigma_H$ now more than halves to about \raisebox{-1ex}
{$\stackrel{\textstyle +3}{\footnotesize -4.5}$}\% ,
reflecting the fact that this
time the maximum and minimum Higgs cross-sections for variable
$\alpha_S$ correspond to $\alpha_S(M_Z^2)=0.1215$ and
$\alpha_S(M_Z^2)=0.116$ respectively. With $\alpha_S(M_Z^2)$ fixed
there is now even a very slight anti-correlation between the
cross-sections.

The extrema in $\sigma_W$, represented by P and R, correspond
roughly to variations in the quark distributions at $x\sim 0.04$,
while the extrema in $\sigma_H$, represented by Q and S,
correspond to variations in the gluon distribution at $x \sim
0.06$ {\em and} $\alpha_S(M_Z^2)$. The values of $x$ sampled at
the Tevatron are an order of magnitude greater than at the LHC.
This, coupled with the fact that it is a proton--antiproton
collider, rather than a proton--proton collider, complicates the
interpretation of the extremes of the cross-sections in terms of
partons.

The up quark distribution for each extrema set with fixed
$\alpha_S(M_Z^2)$ is shown in Fig.~10(a) and the gluon
distribution in Fig.~10(b). The corresponding distributions
obtained when $\alpha_S(M_Z^2)$ is allowed to vary are shown in
Fig.~11. We first consider the cases of the maximum and minimum
$W$ cross-sections, which are insensitive to whether $\alpha_S$ is
left to vary or not. For discussion purposes, let us consider only
the $u$ and $d$ quark flavour contributions. Then the $W$
cross-section is roughly proportional to \be u(x_1)d(x_2)
+d(x_1)u(x_2) + \bar u(x_1)\bar d(x_2) +\bar d(x_1) \bar u(x_2)
\ee where 1 refers to the proton and 2 to the antiproton and
$x_1x_2= M_W^2/s$. Hence the average value of $x_i=0.04$. This is
sufficiently large that there is a distinct difference between the
quark and antiquark distributions, and the contribution to the
cross-section from the quark contribution is the greater. Hence,
one can decrease the cross-section by replacing a quark by its
antiquark at $x=0.05$, or {\it vice versa}. Of course, there is a
fundamental constraint in doing this due to the sum rule for each
valence quark. However, the only real experimental constraint is
from the CCFR $F_3(x,Q^2)$ data, all other structure function data
being insensitive to the distinction between quark and antiquark.
In the optimum global fit most data would like there to be more
quarks at high $x$, while the CCFR $F_3(x,Q^2)$ data would prefer
more valence quarks at $x\leq 0.1$. This leads to a compromise
where for the best fit the CCFR $F_3(x,Q^2)$ data at low $x$ are
undershot. The minimum $\sigma_W$ is therefore achieved mainly by
this exchange of quark for antiquark, which most data are happy
with, and hence the deterioration in $\chi^2$ at P (and P$^*$) is
almost entirely from the description of the CCFR $F_3(x,Q^2)$
data. Hence, both the gluon and quark distribution for P (and
P$^*$) are hardly changed, as seen in Figs.~10 and 11, but $u-\bar
u$ and $d-\bar d$ decrease for $x \sim 0.05$. Going in the other
direction, an increase in $q_V(0.05)$ and the consequent decrease
in the valence quarks at higher $x$ causes a large penalty in
$\chi^2$ and the maximum $\sigma_W$ is achieved in a different
manner. At $x\sim 0.05$ the quark evolves much more slowly than at
$x\sim 0.05$ and the density at $Q^2\sim 10,000\ \GeV^2$ is
determined largely by the input value, and modified by the rate of
evolution. Hence the maximum $\sigma_W$ is achieved by having a
large quark distribution at $x\sim 0.05$ at low $Q^2$ and also by
having an enhanced gluon at $x \sim 0.05$ to increase evolution.
These are displayed in Figs.~10 and 11. The deterioration in
$\chi^2$ then comes mostly from the low $Q^2$ quarks causing an
overshooting of NMC structure function data, but there is also a
contribution due to the enhanced gluon at $x\sim 0.1$ causing it
to be smaller for $x>0.1$ and hence fitting the Tevatron jet data
less well.

The extrema of the Higgs cross-section are also slightly
complicated. It is not possible to simply increase or decrease the
gluon in a range centered on $x\sim 0.05$ because this is
precisely the $x$ region where the majority of the gluon's
momentum is carried, and this total is very well constrained by
the momentum sum rule and the accurate high $x$ quark
determination. Therefore, for fixed $\alpha_S(M_Z^2)$ the change
in $\sigma_H$ is largely reliant on the fact that this total
cross-section actually probes quarks within about an order of
magnitude either side of the central production value of
$x=M_H/\sqrt s$. Hence, as we see from Fig.~10 the maximum
cross-section is obtained from the gluon in set Q$^*$ which is
slightly reduced for $x<0.04$ and more enhanced for $x>0.04$ and
the minimum cross-section is obtained from the gluon in set S$^*$
which is slightly increased for $x<0.04$ and more reduced for
$x>0.04$. In both cases those data sets sensitive to the small $x$
and large $x$ gluon, i.e., HERA structure function data and
Tevatron jet data respectively, are those for which the
description deteriorates. When $\alpha_S(M_Z^2)$ is allowed to go
free it varies by about $\pm 0.003$ and there is a large increase
in the variation of $\sigma_H$. This is not only because $\sigma_H
\propto \alpha_S^2$ but also because the HERA data anti-correlate
$\alpha_S$ and the small $x$ gluon. Therefore, in set Q, for
example, the increased value of $\alpha_S(M_Z^2)$ allows the small
$x$ gluon to get much smaller, and the high $x$ gluon much larger,
compared to set Q$^*$. This compensation between $\alpha_S$ and
the small $x$ gluon also means that HERA data remains well fit,
and it is the jet data (particularly CDF), sensitive to large $x$,
and the large $\alpha_S$-phobic BCDMS data, for which the
description deteriorates. Similar considerations apply to set S as
compared to S$^*$. Here it is the D0 jet data and the small
$\alpha_S$-phobic SLAC and NMC data that are badly fit.

For $\Delta \chi^2 =50$ the Hessian approach gives an uncertainty
of $\pm 1.2 \%$ for $\sigma_W$ and $\pm 3\%$ for $\sigma_H$, at
the Tevatron energy. 
In simplistic terms this is in good
agreement, but a little smaller for the gluon-sensitive Higgs cross-section.
However, in this case we see from Fig.~9 a very marked
asymmetry on the contour plot. For fixed $\alpha_S(M_Z^2)$ the
ellipses are certainly not centered on the best fit values, and
for varying $\alpha_S(M_Z^2)$ we see that $\chi^2$ is clearly
increasing far more rapidly for increases in the predicted $W$
cross-section than for corresponding decreases. Thus, it is clear
that within the framework of this fit, increases of the
cross-section of much more than $3\%$ are completely ruled out,
whereas decreases of the same amount are much more acceptable.
This information would be largely lost in the Hessian approach,
and for these quantities the Lagrange multiplier method does
supply some important additional information.

\section{The ratio of $W^-$ to $W^+$ production at the LHC}

The ratio of the $W^-$ to the $W^+$ production cross-sections at
hadron colliders is a particularly interesting observable. The
measurement is expected to be quite precise (better than $\pm 1\%$
at the LHC, see e.g.~\cite{WPoverWM}), since many of the
experimental uncertainties cancel in the ratio. The uncertainty in
the prediction of the ratio at the LHC can be deduced from the
$\Delta\chi^2$ profile shown in Fig.~12. Taking, as before, the
$\Delta\chi^2 = 50$ measure, we obtain $\Delta(W^-/W^+)=\pm1.3\%$,
and the Hessian approach is in very good agreement with this.
Since the $W^-/W^+$ ratio is sensitive to the ratio of the $d$ and
$u$ quark distributions, it is not surprising that the increase in
$\chi^2$ is almost entirely due to the NMC $F_2(n)/F_2(p)$
data~\cite{NMC}.

A detailed discussion of the $W^-/W^+$ ratio may be found in
Ref.~\cite{MRSTWZ}. Consider, for instance, the ratio as a
function of the $W$ rapidity $y$
\be \label{eq:ratio_func_y}
\frac{d\sigma/dy(W^-)}{d\sigma/dy(W^+)}\ \simeq\
\frac{d(x_1)\bar{u}(x_2)}{u(x_1)\bar{d}(x_2)}\ \simeq\
\frac{d(x_1)}{u(x_1)}, \ee
where $x_1 = M_We^y/\sqrt{s} = 0.0057e^y$ at the LHC. In
Eq.~(\ref{eq:ratio_func_y}) we have ignored, for simplicity, the
contributions involving strange and heavier quarks. Thus a
measurement of the ratio at large $y$ would provide a direct
determination of $d/u$ at large $x$. For example, for $y\simeq 4$,
we measure $d/u$ at $x\sim0.3$ at the LHC. Of course, it is the
decay lepton rapidity that is measured, rather than the parent $W$
rapidity, and so the ratio in a given rapidity bin will have a
greater uncertainty than that for $\sigma(W^-)/\sigma(W^+)$.

\section{The moments of the ($u$--$d$) distribution}

The parton distribution functions of the nucleon are fundamental
quantities that should, in principle, be calculable from first
principles in QCD. In particular, the $x$ {\it moments} of parton
distributions at a given scale $Q^2$ are related, by the operator
product expansion, to a product
 of perturbatively calculable Wilson coefficients and non-perturbative matrix
elements of quark and gluon operators. The latter can be computed
using lattice QCD and, indeed, in recent years the precision of
the lattice calculations has improved significantly. Although in
principle the lattice results can be related to moments of
physical structure functions, in practice it is more efficient to
use parton distributions determined in a global fit to represent
the physical `data'. Comparisons of recent lattice moment
calculations with the predictions of earlier MRS parton
distributions are encouraging, see for example
\cite{LHPC-SESAM,QCDSF}.

In order to quantify the agreement between the lattice
calculations and the parton distribution predictions it is
obviously important to know the uncertainties in the latter. It is
straightforward to apply the Lagrange multiplier method used in
previous sections to determine the uncertainties in observable
cross-sections to the moments of parton distributions.

To avoid contamination from gluon contributions, the lattice
calculations focus on the moments of non-singlet quark operators.
For example, lattice results are available for the first three
moments of the combination $u-d$, i.e.,
\be \label{eq:M_N} M_N^{u-d}(Q^2) = \int_0^1 dx\; x^{N-1}\;
[u(x,Q^2)-d(x,Q^2)] \ee
with $N=2,3,4$. The predictions of the MRST2001 set (at $Q^2 = 4\
{\rm GeV^2}$) for these moments are given in Table~1.

The $\Delta \chi^2$ contour plot for the (percentage) variation of
the second and third moments about their predicted values is shown
in Fig.~13. We again show the $\Delta \chi^2 = 50$ and $100$
contours corresponding to the fixed $\alpha_S$ analysis, but there
is evidently little difference between the fixed and variable
coupling results in this case.

As expected, there is a strong positive correlation between the
two moments. Using the $\Delta \chi^2 = 50$, varying $\alpha_S$
criterion for defining a conservative error, we obtain errors of
 $\pm 4.2\%$, $\pm 4.8\%$ and $\pm 5.0\%$  for the second, third and fourth
moments respectively.
 The corresponding predictions for the errors on the moments are also given in Table~1. The increasing
relative error with increasing moment is to be expected -- higher
moments probe the $x \to 1$ region where there are fewer DIS
structure function data. Again we notice that there is a small
asymmetry in the contours -- the increase in $\chi^2$ when both
moments increase being less severe than when both moments
decrease.
\begin{table}[h]\begin{center}
\begin{tabular}{|c|c|c|}\hline
$N$          & $M_N^{u-d}(4\; {\rm GeV^2})$ & \% error \\ \hline
2           & 0.1655(70)      &  4.2  \\
3        &    0.0544(26)    & 4.8 \\
4    &   0.0232(12)     &  5.0   \\
\hline
\end{tabular}\caption{The moments and their errors of the ($u$--$d$)
distribution, Eq.~(\ref{eq:M_N}), predicted at $Q^2=4\ \GeV^2$ using
MRST2001 partons~\cite{MRST2001}.}
\end{center}\end{table}

 The uncertainties on the moments using the Lagrange multiplier
method with a {\it fixed} $\alpha_S$ are slightly smaller: $\pm
4.1\%$, $\pm 4.3\%$ and $\pm 4.7\%$ for the second, third and
fourth
 moments respectively. These results are in excellent agreement with the
(fixed $\alpha_S$) Hessian approach, where the corresponding
errors are
 $\pm 3.9\%$, $\pm 4.3\%$ and $\pm 4.6\%$.

Since, as we have already seen in Section~4, the $u$ quark at high
$x$ is far more constrained than the $d$ quark, the allowed
variation in these moments is mainly due to variations in the
$d_V$ distribution. The minimum extremum (H in Fig.~13) of the
moments is therefore due to the largest allowed $d_V$ distribution
at high $x$ and arises from a similar set of partons to those for
the maximum $F_2^{CC}(e^+ p)$. Thus, as in this previous case, it
is mainly the comparison to the BCDMS $F_2(e d)$ measurements
which causes the deterioration in the quality of the fit. The
maximum of the moments (G in Fig.~13) corresponds roughly to the
minimum $d_V$ distribution at high $x$ and it is largely the fit
to the $F_2(n)/F_2(p)$ ratio that breaks down.

For a number of years, lattice QCD has been used to calculate the
moments of nucleon structure functions from first principles. The
most recent comprehensive results are from the LHPC-SESAM
\cite{LHPC-SESAM} and QCDSF \cite{QCDSF} collaborations. Although
the comparisons with experiment (via parton distributions obtained
from global fits) are encouraging, there are still many problems
to be overcome, for example finite lattice spacing and volume
effects, renormalization and mixing of operators, unquenching and
chiral extrapolation to physical quark masses.  A comparison with
the recent lattice results \cite{LHPC-SESAM,QCDSF}  and the above
MRST2001 moment predictions reveals that (a)~the errors in the
latter are at present significantly smaller than in the former,
especially for the higher moments, and (b)~the lattice results for
the moments are systematically higher. The explanation appears to
be that the linear chiral extrapolation used in the lattice
determinations is not valid -- non-perturbative long-distance
effects in the nucleon gives rise to nonlinear, non-analytic
dependence on $m_q$ \cite{Detmold1}--\cite{ChenJi} which is
particularly important at small $m_q$. In the most recent analyses
(see for example the comprehensive study in \cite{Thomas2}), the
experimental (i.e., pdf) values for the moments are used to
constrain {\it a priori} unknown non-perturbative parameters which
enter in the non-analytic terms in the chiral extrapolation
formula. It will be interesting to investigate the effect of using
the  new MRST2001 moment predictions and errors in such studies.

\section{Comparison between different central parton sets}

So far in this paper we have investigated the uncertainty on
physical quantities coming from the experimental error of the
measurements used to determine the parton distributions. We have
discussed both the Hessian and Lagrange Multiplier approaches,
concluding that the latter is in principle preferable, but
recognizing the practical advantages of the former. We have
compared the results each provide for the uncertainties using the
$\Delta\chi^2 =50$ criterion, noting that they are generally in
good agreement. The Hessian approach does tend to give slightly
smaller uncertainties for the quantities sensitive to the least
well-determined partons, i.e., $\sigma_H$  which is sensitive to
the gluon distribution and $F_2^{CC}(e^+ p)$ which is sensitive to
the high-$x$ down quark distribution. This is probably partly due
to the neglected effect of the not entirely redundant parameters,
and partly due to errors associated with those eigenvectors which
do not respect the quadratic approximation for $\Delta \chi^2$ too
well, which indeed are mainly concerned with the gluon and high
$x$ down quark. However, the discrepancy is quite small, and we
judge that we can trust the Hessian approach, at least for $\Delta
\chi^2$ in the region of 50 or less, to give quantitative results.
Hence, for fixed $\alpha_S(M_Z^2)=0.119$, we have made available
30 parton sets corresponding to the 15 different eigenvector
directions in the space of variation of parton parameters away
from their values at the minimum $\chi^2$ of the global fit, each
set corresponding to an increase in $\chi^2$ of 50. These can
easily be used to obtain the error on any physical quantity, as
outlined in Section~2. We have also made available various parton
sets with fixed and varying $\alpha_S(M_Z^2)$ corresponding to
extreme variations in the predictions for various important
cross-sections and other relevant observables.

We note that the uncertainties obtained due to the errors on the
experimental data are generally very small, of the order of
$1-5\%$, except for quantities sensitive to the high-$x$ down
quark and gluon, where they can approach $10\%$. However, in all
of this we have implicitly assumed that the theoretical procedure
is precisely compatible with the data used, we have not considered
the uncertainties due to (i)~the data sets chosen, (ii)~the choice
of starting parameterizations, (iii)~the heavy target corrections,
{\it etc}. In practice this is far from true, as discussed in the
Introduction. In this final section we acknowledge this to some
extent and investigate qualitatively the impact of the initial
assumptions going into the fit on the uncertainty on some
quantities. In order to do this we first perform a slightly
updated fit of our own (which includes minor modifications in
terms of parameterization and the treatment of errors and data
sets) so as to produce the best set of up-to-date partons. This
was partially inspired by the question of why CTEQ6 \cite{CTEQ6}
gives a much better fit to the Tevatron jet data than MRST2001,
but also by the availability of new ZEUS data. We call the new set
MRST2002 partons.\footnote{The MRST2002 parton set can be found at
http://durpdg.dur.ac.uk/hepdata/mrs$\,$.}

\subsection{CTEQ6, MRST2001 and a new parton set (MRST2002)}

We found that we can improve the fits to jets within the global
fit by a couple of modifications. In order to obtain the best
global fit with partons input at $Q_0^2=1\ \GeV^2$ we had previously
found that we needed a parameterization which allows the gluon to
go negative at small $x$. Hence we used
\begin{equation}
xg(x,Q_0^2)=A_g(1-x)^{\eta_g}(1+\epsilon_gx^{0.5}+\gamma_g x)x^{\delta_g}
-A_-(1-x)^{\eta_-}x^{-\delta_-},
\label{eq:gluon1}
\end{equation}
where $A_- \sim 0.2$, $\delta_- \sim 0.3$ and $\eta_-$ was fixed
at $\sim 10$, so as not to affect the high $x$ distribution.
Unexpectedly, allowing $\eta_-$ to vary to $\sim 25$ resulted in a
slight improvement in the fit to Tevatron jets. We also modified
our treatment of the errors for the Drell--Yan data
\cite{DrellYan}. The fit to these data actually competes with that
to the jets, and using only statistical errors, as in our previous
studies (the systematic errors being defined a little vaguely),
over-emphasizes the effect of the Drell--Yan measurements. Adding
$5\%$ systematic errors in quadrature to the statistical errors
(which is probably the best approach \cite{DrellYan}) also
improves the fit to the jet data. Both these modifications appear
appropriate and are implemented in our updated set. Also included
in the new analysis is the new ZEUS high-$Q^2$
data~\cite{ZEUSnew}, which has little effect on the partons. The
only significant change in the MRST2002 partons, compared to
MRST2001 partons~\cite{MRST2001}, is an increase in the gluon at
high $x$, which we show in Fig.~14. The fit to the Tevatron jet
data now has $\chi^2=154/113$ compared to $\chi^2=170/113$ for
MRST2001, and the fit to The Drell--Yan data with $5\%$ systematic
errors has $\chi^2=187/136$. The quality of fit for all other data
sets is almost identical to that for the MRST2001 partons.

The CTEQ6 partons are very similar to the MRST2001 (and MRST2002)
partons in most aspects. However, in this CTEQ
analysis~\cite{CTEQ6} a number of different choices are made about
the way in which the fit is implemented, which leads mainly to a
significantly different gluon distribution. These differences are:
the development of a different type of parameterization for the
partons, which allows for a different shape at very high $x$; CTEQ
omit data below $Q^2 = 4\ \GeV^2$, compared to our choice of $Q^2
= 2\ \GeV^2$; they do not fit to some data sets used in
\cite{MRST2001}, i.e., they omit SLAC and one H1 high-$Q^2$ set of
$F_2$ data; they use $10\%$ systematic errors (in quadrature) for
Drell--Yan data; moreover, CTEQ have a positive-definite small-$x$
gluon at their starting scale of $Q_0^2= 1.69\ \GeV^2$. They also
use a massless charm prescription and there are various other
minor differences as compared with MRST.\footnote{The way in which
these different assumptions lead to an improved fit to the
Tevatron jet data is outlined in \cite{krakow}.}

The CTEQ6 gluon is also shown in Fig.~14. Clearly MRST2002 has a
similar high-$x$ gluon to CTEQ6, both being larger than MRST2001.
However, the MRST gluons are different from the CTEQ6 gluon at
smaller $x$ due to their freedom to have a negative input
distribution, and due to slight differences in the choice of data
sets fitted. The different assumptions made in obtaining the CTEQ
partons, although they improve the quality of the jet fit, do not
lead to the best fit when including the data sets omitted by CTEQ
and the fit is not good at all for data with $Q^2 < 4\ \GeV^2$.
Hence, within the context of trying to obtain as inclusive a
global fit as possible using NLO QCD, we take MRST2002 to be the
best set of parton distributions.

\subsection{Comparison of predictions for $\sigma_W$ and for
$\sigma_H$}

The predictions for $W$ and Higgs cross-sections using the
different partons are shown in Fig.~15. Since MRST2002 only
differs from MRST2001 in the high $x$ gluon, to which these
cross-sections are insensitive, the predictions for MRST2002 are
very similar to those of MRST2001. (Hence our decision to keep
MRST2001 partons as the base set for this paper). However, the
corresponding predictions obtained using the CTEQ6 partons are
quite different. At the LHC the prediction for $\sigma_W$ is
similar, but $\sigma_H$ is towards the top of our (qualitative)
$95\%$ confidence level. From Fig.~14 this is clearly due to the
larger gluon in the $x \sim 0.005$ region, which is due to the
positive definite input for the CTEQ6 gluon. At the Tevatron the
discrepancy between CTEQ6 and MRST is even larger. The CTEQ6
predictions for both $\sigma_W$ and $\sigma_H$ are effectively
completely outside our expectations. The reason for the small
prediction of $\sigma_H$ is evident from Fig.~14---the CTEQ6 gluon
is considerably smaller in the region of $x=0.1$. This, in turn,
is then responsible for a slower evolution of the quarks, making
them smaller at high $Q^2$ and hence making $\sigma_W$ smaller.
Presumably the difference comes about because CTEQ6 use a more
restricted form of the gluon and omit one H1 data set and $Q^2\leq
4\ \GeV^2$ data which prefer larger $dF_2(x,Q^2)/d\ln Q^2$.
Whatever the precise reasons for the discrepancies, it is clear
that different choices for the overall framework of the global fit
can completely outweigh the uncertainties due to errors on the
data actually chosen to go into the fit. It would be easy to
illustrate similar types of discrepancy comparing to other
alternative sets of partons---in particular, due to the absence of
the Tevatron jets in the fits, many of the parton sets in
\cite{Botje}--\cite{ZEUSfit} have rather smaller gluons at large
$x$, and would have different predictions for various quantities
sensitive to the high-$x$ gluon.

\begin{table}
\begin{center}
\renewcommand{\arraystretch}{1.4}
\setlength\tabcolsep{5pt}
\begin{tabular}{lll}
\hline\noalign{\smallskip}
Group & variation & $\alpha_S(M_Z^2)$ \\
         & of $\chi^2$ & \\
\noalign{\smallskip}
\hline
\noalign{\smallskip}
CTEQ6 & $\Delta \chi^2 = 100$ & $0.1165
  \pm 0.0065({\rm exp}) $ \\
  ZEUS  & $\Delta \chi_{\rm eff}^2 = 50$ &$ 0.1166
  \pm 0.0049({\rm exp})\pm 0.0018({\rm model})$ $\pm 0.004({\rm theory}) $ \\
  MRST02 & $\Delta \chi^2 = 20$
  &$ 0.1195 \pm 0.002({\rm exp})\pm 0.003({\rm theory})  $ \\
  MRST01 & $\Delta \chi^2 = 20$
  &$ 0.1190 \pm 0.002({\rm exp})\pm 0.003({\rm theory})  $ \\

  H1    & $\Delta \chi^2 = 1$ &
  $0.115 \pm 0.0017({\rm exp})~^{+0.0009}_{-0.0005}~({\rm model})$
  $\pm 0.005({\rm theory})$ \\
  Alekhin & $\Delta \chi^2 = 1$ & $0.1171
  \pm 0.0015({\rm exp}) \pm 0.0033({\rm theory})$ \\
 GKK & $\Delta \chi_{\rm eff}^2 = 1$  & $0.112 \pm 0.001({\rm exp})$ \\
\hline
\end{tabular}
\vspace{0.3cm} \caption{Values of $\alpha_S(M_Z^2)$ and its error
from different NLO QCD fits.}
\end{center}
\label{Tab1}
\end{table}

\subsection{Comparison of predictions for $\alpha_S(M_Z^2)$}

We also find a large variation in the values of $\alpha_S(M_Z^2)$
extracted from the fits of the different collaborations:
CTEQ6~\cite{CTEQ6}, ZEUS~\cite{ZEUSfit}, MRST2001~\cite{MRST2001},
H1~\cite{H1Krakow}, Alekhin~\cite{Alekhin} and Giele et~al.
(GKK)~\cite{Giele}. The resulting values of $\alpha_S(M_Z^2)$ are
listed in Table~2, together with the determination of this work
(MRST2002), in order of decreasing tolerance
($\sqrt{\Delta\chi^2}$), which is reflected in the size of the
corresponding experimental error. Not all are presented as
determinations of $\alpha_S(M_Z^2)$, but all are extracted using
the same criteria as for the uncertainty on partons in the
respective fit, and hence should be as reliable. Clearly there is
a very large variation, with some very low values. The
uncertainties due to experimental errors are determined in
different fashions in each case, and a summary can be found in
\cite{stats}. We use $\Delta \chi^2_{\rm eff}$ for the ZEUS
determination~\cite{ZEUSfit}, because they use the offset method
for determining uncertainties which for $\Delta \chi^2 =1$ gives a
larger uncertainty than the more common Hessian method. ZEUS
estimate that this is equivalent to $\Delta \chi^2 \approx 50$ if
they were to use the same treatment of errors as CTEQ. We also use
$\Delta \chi^2_{\rm eff}$ for the GKK value \cite{Giele}, because
the uncertainties are obtained using confidence limits, but the
error quoted corresponds to the one sigma usually associated with
$\Delta \chi^2=1$.

The model errors incorporate such effects as the heavy quark
prescription and masses, parameterizations, changes in the
starting scale of evolution {\it etc}. The theory error is often
determined by variation of renormalization and factorization
scales, though MRST use an estimate appealing to current knowledge
of NNLO and resummations, which we feel is more reliable. Since
each fit is centered on NLO QCD with scales equal to $Q^2$, the
``theory errors'' are very strongly correlated, and cannot
therefore be responsible for the differences. These discrepancies
are undoubtedly due to the assumptions going into the fits, mainly
on which data sets are included and which cuts on $Q^2$ and $W^2$
are used.

MRST, who obtain the largest value of $\alpha_S(M_Z^2)$, use the
widest range of data sets and also the least conservative
cuts.\footnote{The slightly different treatment in this work
(MRST2002) leads to a marginal raising of $\alpha_S(M_Z^2)$ as
compared to MRST2001~\cite{MRST2001}, as seen in Table~2. We still
use $\Delta \chi^2 =20$ for
our one-sigma uncertainty, since if $\Delta \chi^2 =50$ corresponds to
90\% confidence level, or 1.65 sigma, simple scaling implies that one sigma 
corresponds to $\Delta \chi^2 =50/(1.65)^2$, 
i.e. $\Delta \chi^2 =20$ to a good approximation.} CTEQ use only a
slightly smaller number of data sets but also cut data below
$Q^2=4\ \GeV^2$, as described previously. They also use a
definition of the NLO coupling which truncates the solution of the
renormalization group equation, whereas most other groups use the
full solution of the NLO equation. Both approaches are equally
correct, but the truncation of the solution leads to a slightly
higher value of $\alpha_S(Q^2)$ at scales below $M_Z^2$, for the
same value of $\alpha_S(M_Z^2)$, than the other method, and thus
tends to yield a lower $\alpha_S(M_Z^2)$. CTEQ also have a very
conservative estimate of the error, though it is meant to be
somewhat more than a one-sigma error. ZEUS and Alekhin use a
similar selection of data sets, i.e., HERA DIS data (only ZEUS
data in the former case) and a number of fixed target DIS data
sets. Hence, it is unsurprising that they obtain similar central
values of $\alpha_S(M_Z^2)$, with respective errors which are
easily explained by their choices of $\Delta \chi^2$. H1 and GKK
both use a small number of sets of data: the former collaboration
uses H1 DIS data \cite{H1Krakow,H1A} and BCDMS fixed-target proton
DIS data \cite {BCDMSp}, while GKK use older H1 DIS data
\cite{H194} together with BCDMS and E665 \cite{E665} fixed-target
proton DIS data. Both determinations are heavily influenced by the
BCDMS proton data set which prefers rather small\footnote{Recall
the determination $\alpha_S(M_Z^2)=0.113\pm0.005$ from BCDMS data
alone~\cite{VM}.} $\alpha_S(M_Z^2)$, and this feeds into the final
values. Also, both are strict in their statistical interpretation,
obtaining small uncertainties, even with relatively small data
samples. Finally we note that only CTEQ and MRST include the
Tevatron jet data in their analyses. This is relevant because of
the $\alpha_S$--gluon correlation.

\subsection{Final comment}

From the discussion of the previous two subsections, it is clear
that different ideas about the best way to perform a NLO fit can
lead to a wide variation in both the central values and the errors
of $\alpha_S(M_Z^2)$ as well as in predictions for physical
quantities such as $\sigma_W$ and $\sigma_H$. The fact that the
various `NLO' fits can yield such different outputs is disturbing,
and is indicative of the uncertainty arising from theoretical
assumptions. Indeed, we have always believed that `theory', rather
than experiment, will provide the dominant source of error
\cite{stats}. We have already produced approximate NNLO parton
distributions and predictions \cite{MRSTNNLO} (based on the
approximate splitting functions \cite{NNLOsplit} obtained from the
known NNLO moments \cite{NNLOmoms}), and find, for example, that
the NNLO $W$ cross-section at the Tevatron is $4\%$ higher than at
NLO, and believe that this result is reliable. This change is
somewhat larger than the uncertainty due to experimental errors
shown in Fig.~9. Moreover, $W$ production is likely to be subject
to smaller theoretical uncertainty than many other
observables---particularly those directly related to the gluon.
Our estimates for the uncertainty in $F_L(x,Q^2)$ at small $x$ are
$10\%$ or more even at $Q^2=10,000\ \GeV^2$, and significantly
larger at lower $Q^2$, for example.  Hence, an understanding of
theoretical uncertainties is clearly a priority at present, and a
preliminary attempt at this will be the subject of a future
publication \cite{cuts}.

\section*{Acknowledgements}

We would like to thank Mandy Cooper-Sarkar, Vladimir Chekelian,
Dan Stump and Wu-ki Tung for useful discussions. RST would like to thank
the Royal Society for the award of a University Research Fellowship. RGR
would like to thank the Leverhulme Trust for the award of an Emeritus
Fellowship. The IPPP gratefully acknowledges financial support from the UK
Particle Physics and Astronomy Research Council.

\newpage

\begin{figure}[htbp]
\epsfig{figure=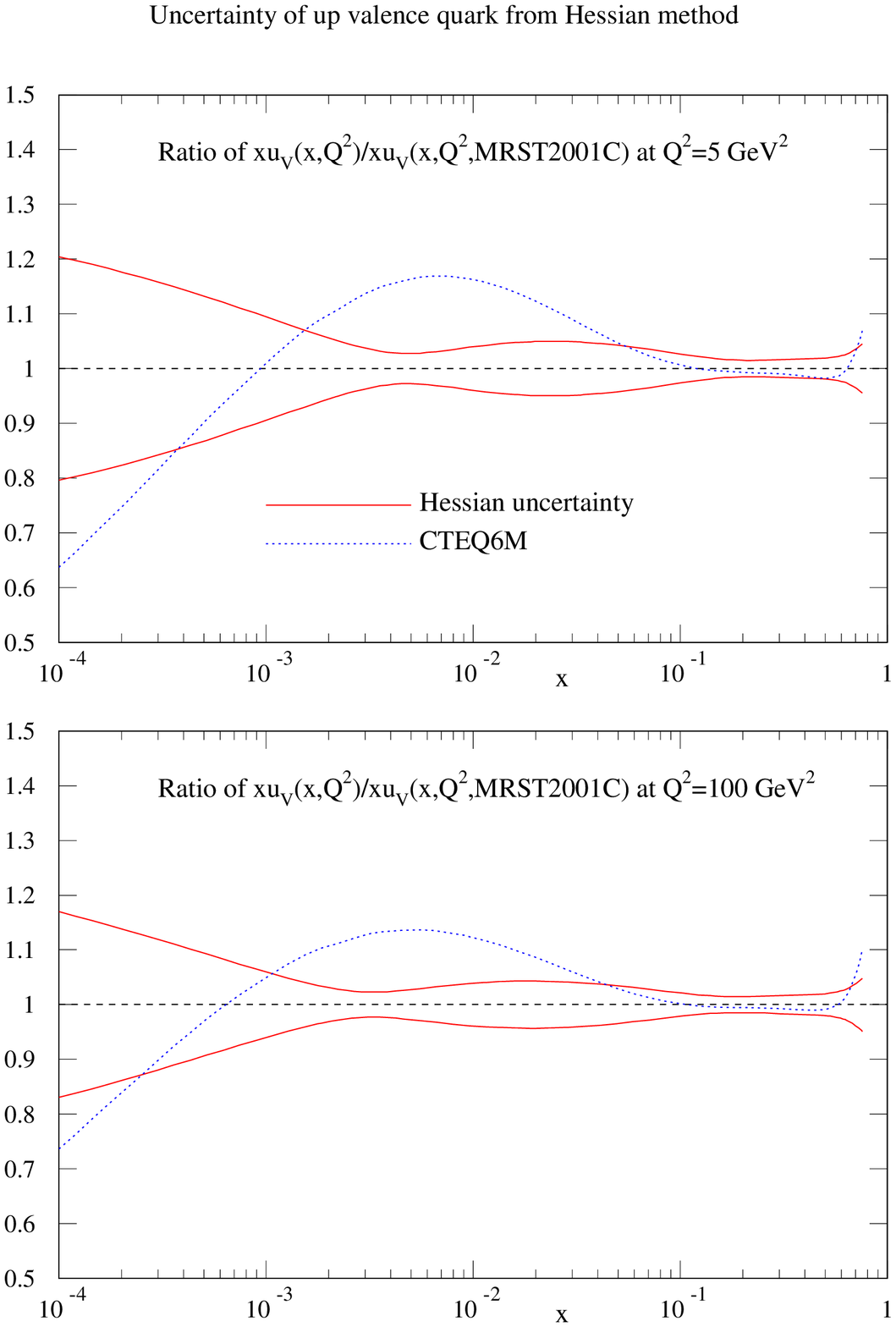,height=8in}
\vspace{-1.75cm}
 \caption{The uncertainty on $u_V(x,Q^2)$ at $Q^2= 5\ \GeV^2$
and $100\ \GeV^2$ obtained using the Hessian approach with $\Delta
\chi^2=50$. Also shown is the CTEQ6M distribution. The
uncertainties are shown relative to the MRST2001 set of
partons~\cite{MRST2001}; the label C is explained in footnote~5.}
\label{Fig1}
\end{figure}

\begin{figure}[htbp]
\epsfig{figure=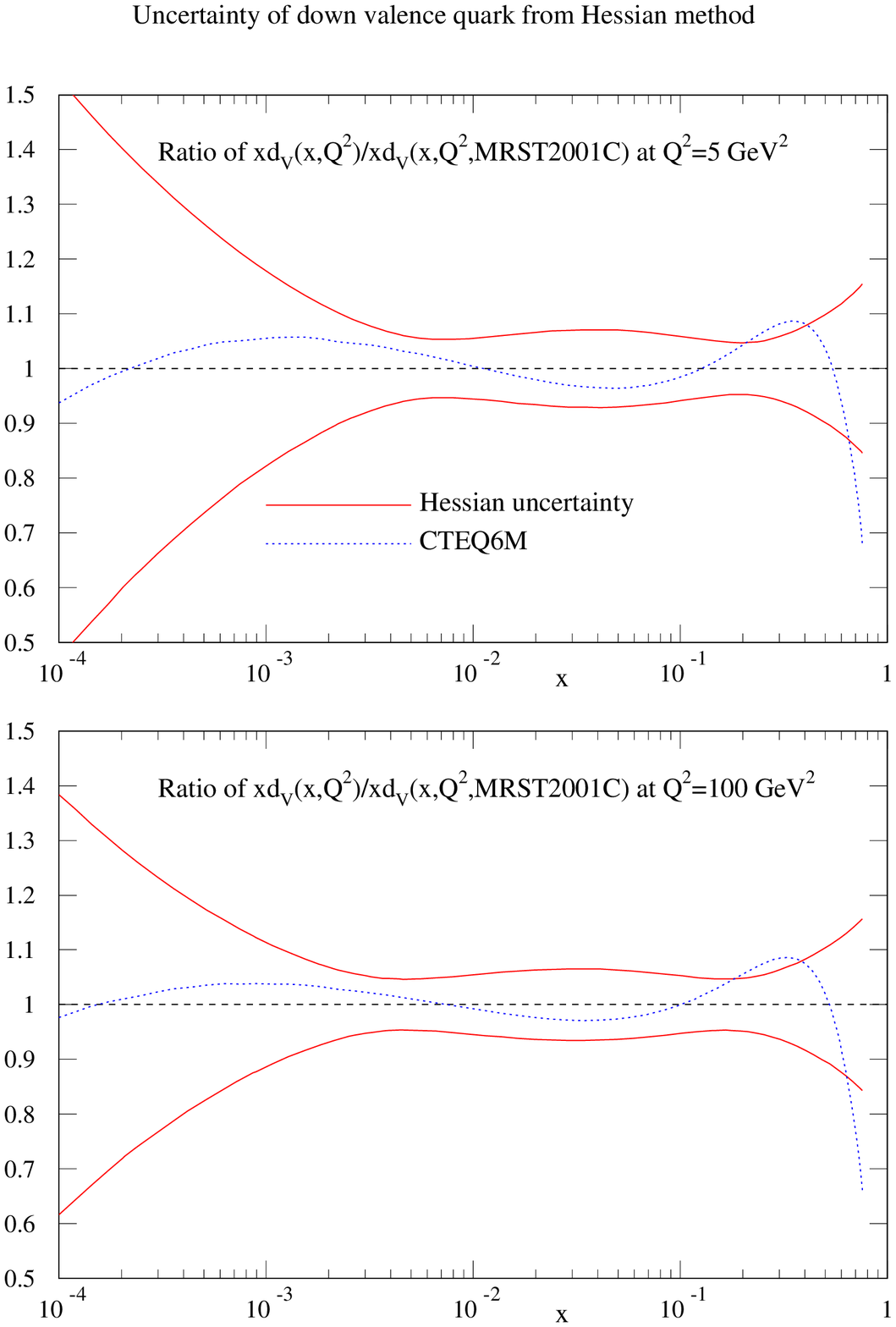,height=8in}
\vspace{-1.75cm}
 \caption{The uncertainty on $d_V(x,Q^2)$ at  $Q^2= 2\ \GeV^2$
and $100\ \GeV^2$ obtained using the Hessian approach with $\Delta
\chi^2=50$. Also shown is the CTEQ6M distribution. The
uncertainties are shown relative to the MRST2001 set of
partons~\cite{MRST2001}; the label C is explained in footnote~5.}
\label{Fig2}
\end{figure}

\begin{figure}[htbp]
\epsfig{figure=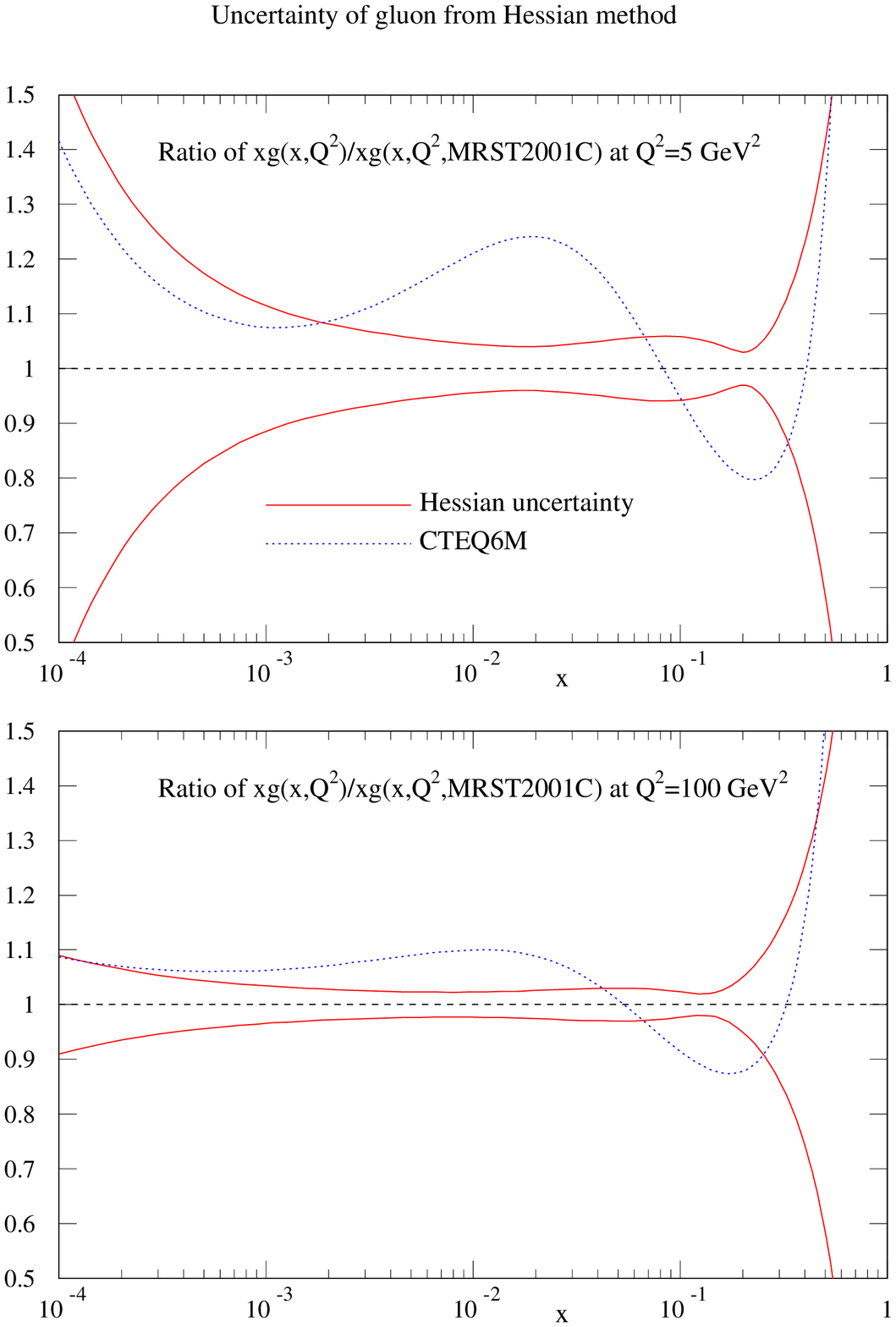,height=8in}
\vspace{-1.75cm}
 \caption{The uncertainty on $g(x,Q^2)$ at $Q^2= 5\ \GeV^2$
and $100\ \GeV^2$ obtained using the Hessian approach with $\Delta
\chi^2=50$. Also shown is the CTEQ6M distribution. The
uncertainties are shown relative to the MRST2001 set of
partons~\cite{MRST2001}; the label C is explained in footnote~5.}
\label{Fig3}
\end{figure}

\begin{figure}[htbp]
\epsfig{figure=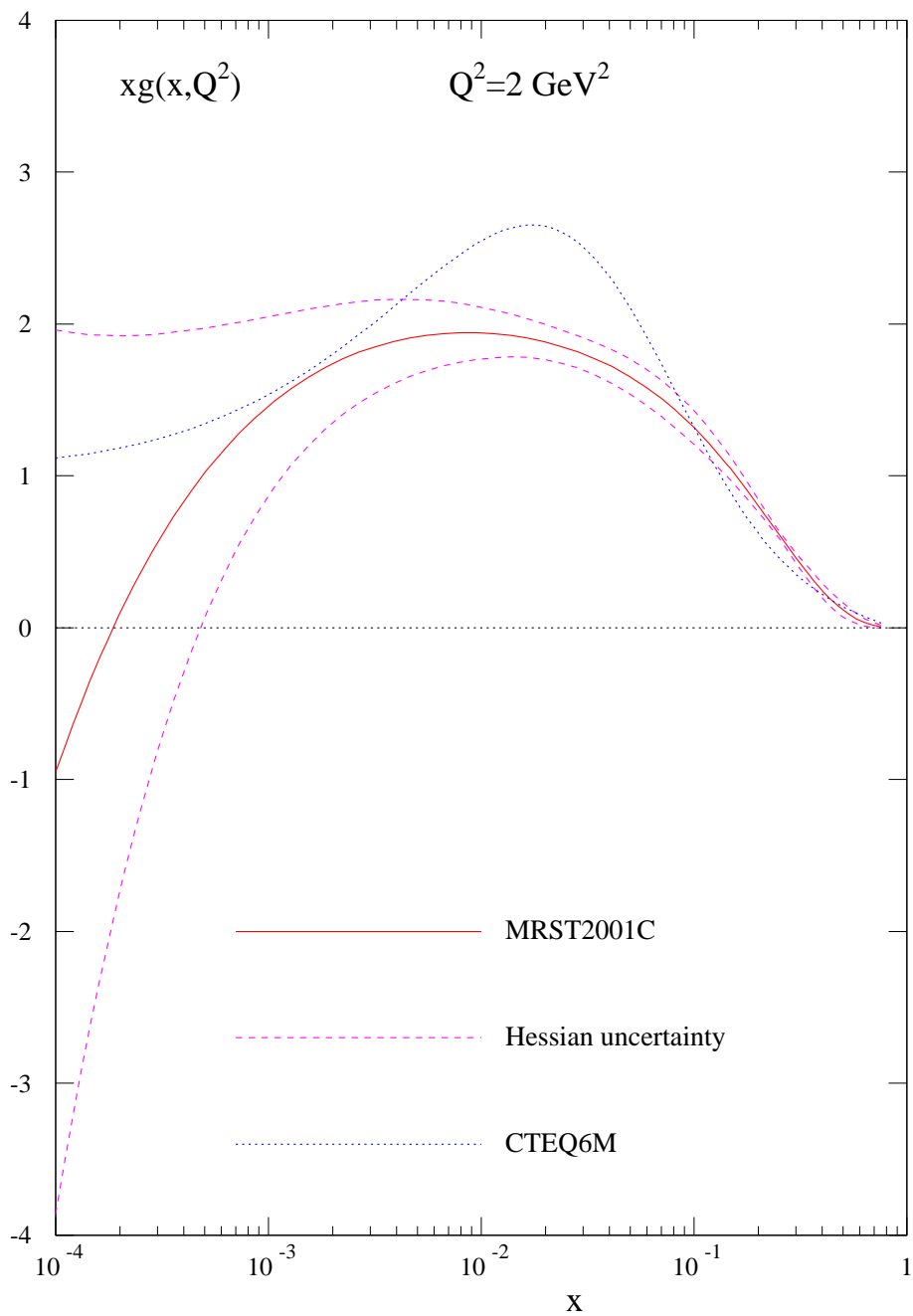,height=8in}
\vspace{-1.75cm}
 \caption{The uncertainty on $g(x,Q^2)$ at $Q^2= 2\ \GeV^2$
obtained using the Hessian approach with $\Delta \chi^2=50$. Also
shown is the CTEQ6M distribution.} \label{Fig4}
\end{figure}

\begin{figure}[htbp]
\epsfig{figure=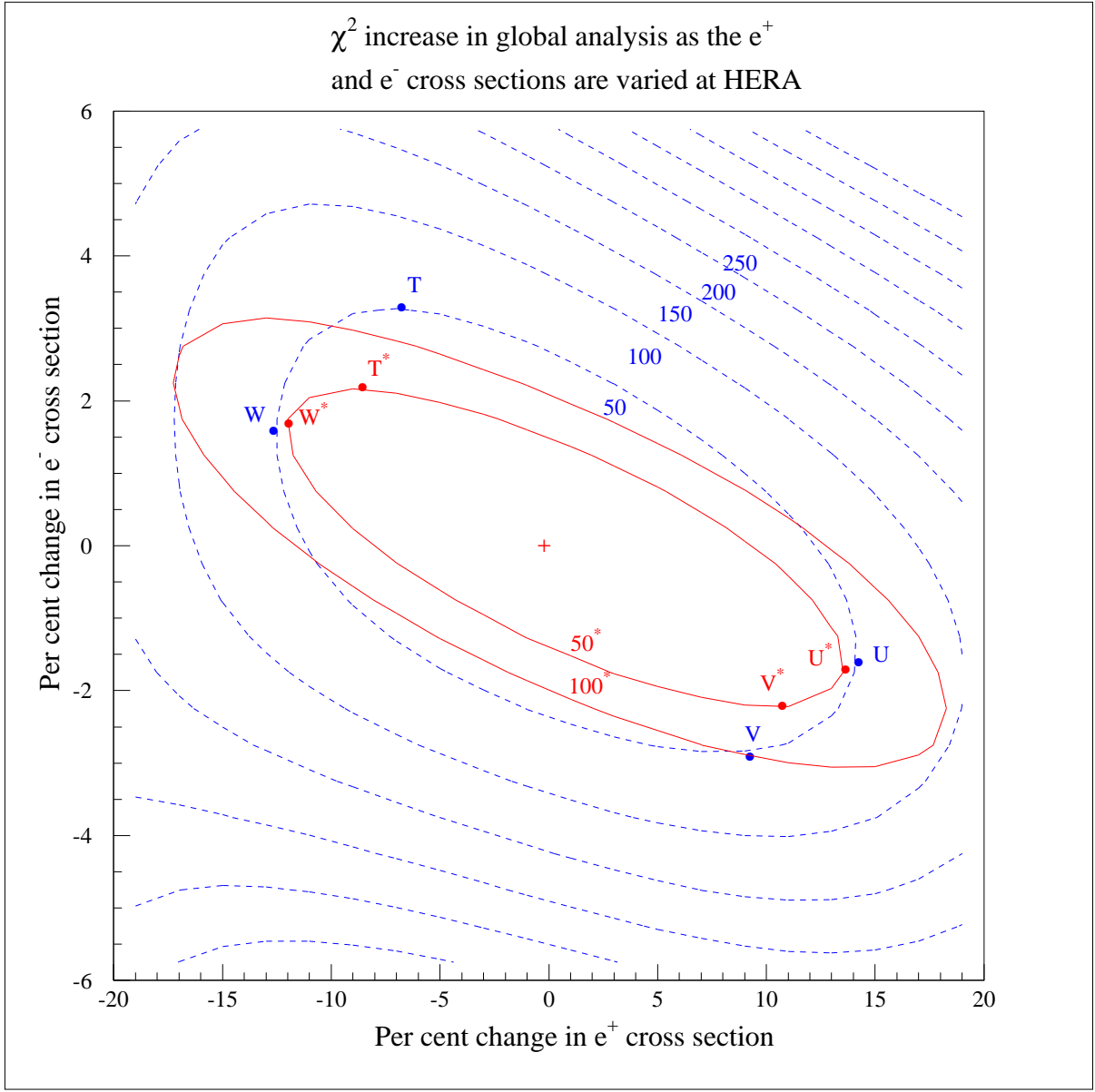,height=8in}
\vspace{-1.75cm}
 \caption{$\Delta\chi^2=50,100,\dots$ contours, where $\Delta\chi^2$ is the
increase in $\chi^2$ from the global MRST2001 minimum, obtained by performing
new global fits with $F_2^{CC}(e^\pm p)$ fixed at values in the neighbourhood
of their value in unconstrained MRST2001 fit. The $\Delta\chi^2=50$ contour
is taken to represent the errors on $F_2^{CC}(e^\pm p)$ (arising from the
experimental errors on the data used in the global fit). The extrema sets of
partons (T,U,$\dots$) are discussed in the text. The dashed
contours are obtained if $\alpha_S(M_Z^2)$ is allowed to vary. The
superimposed solid  $\Delta\chi^2=50,100$ contours are obtained if
$\alpha_S(M_Z^2)$ is fixed at 0.119.}
\label{Fig5}
\end{figure}

\newpage

\begin{figure}[htbp]
\hspace{-2.5cm}
\includegraphics[width=9.5cm]{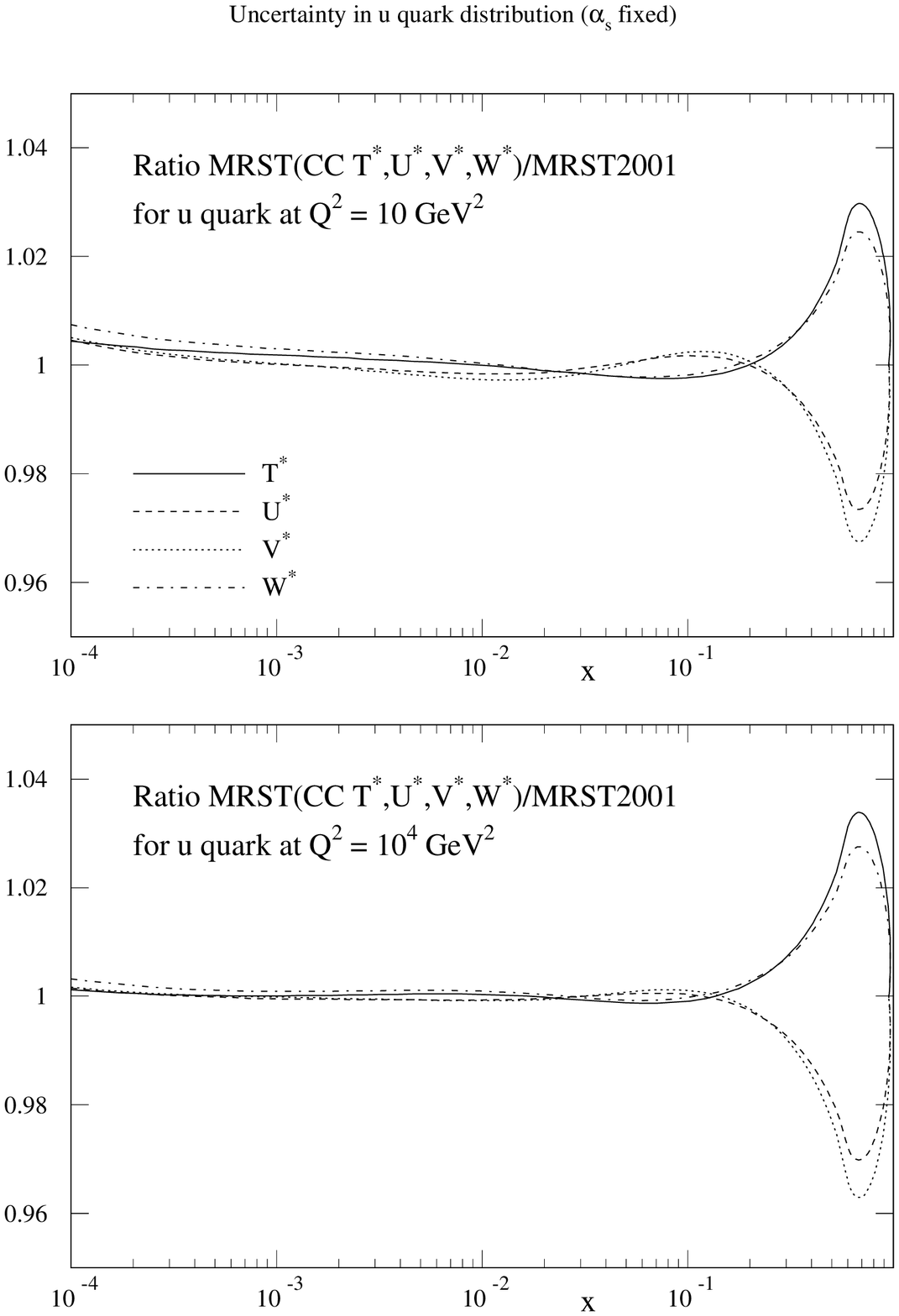}
\hspace{0.0cm}
\includegraphics[width=9.5cm]{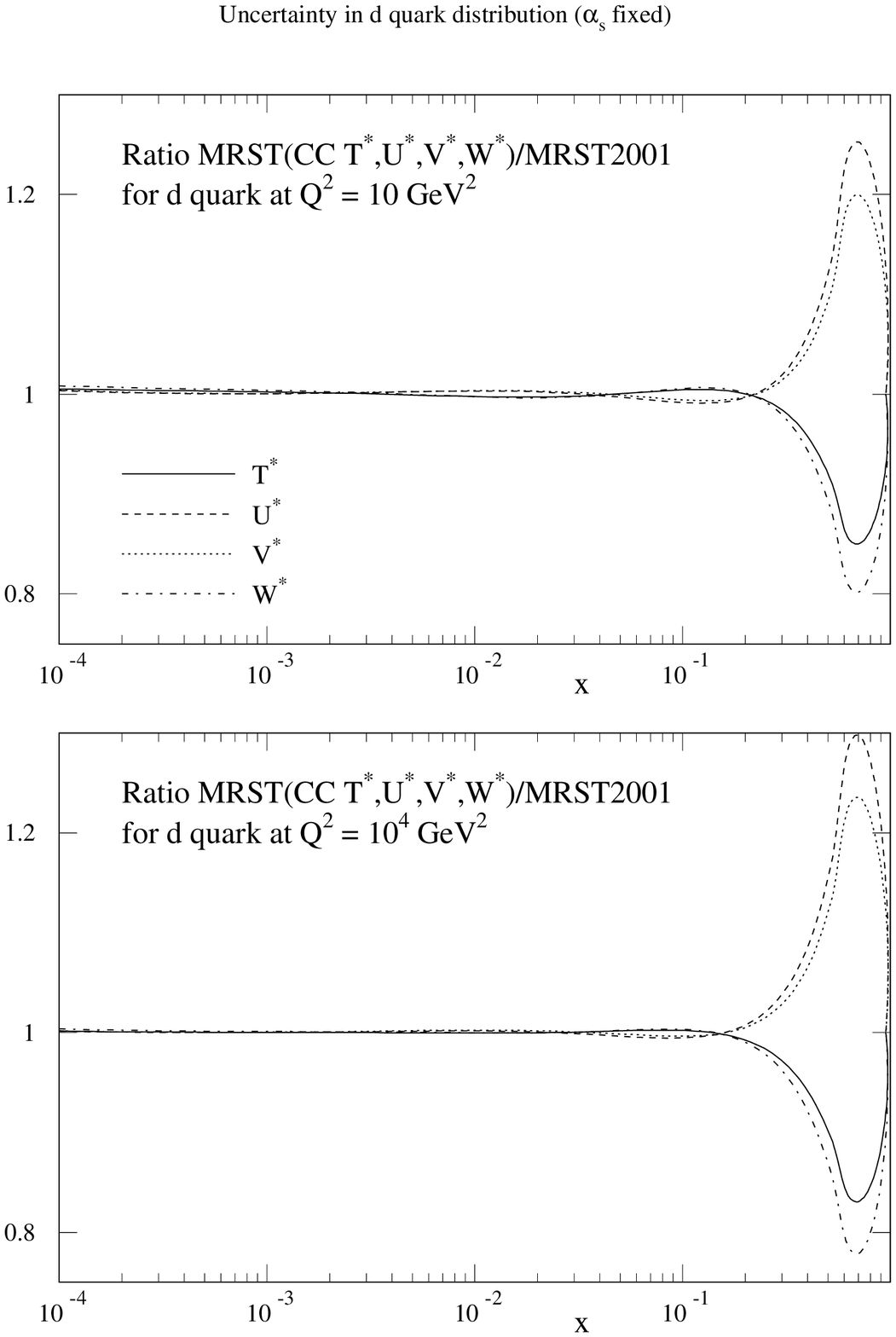}
\caption{The $u$ and $d$ quark distributions (at $Q^2=10$ and
$10^4\ \GeV^2$) of the extrema fits which lie on the
$\Delta\chi^2=50$ contour of Fig.~5 for fixed
$\alpha_S(M_Z^2)=0.119$.} \label{Fig6}
\end{figure}

\newpage

\begin{figure}[htbp]
\epsfig{figure=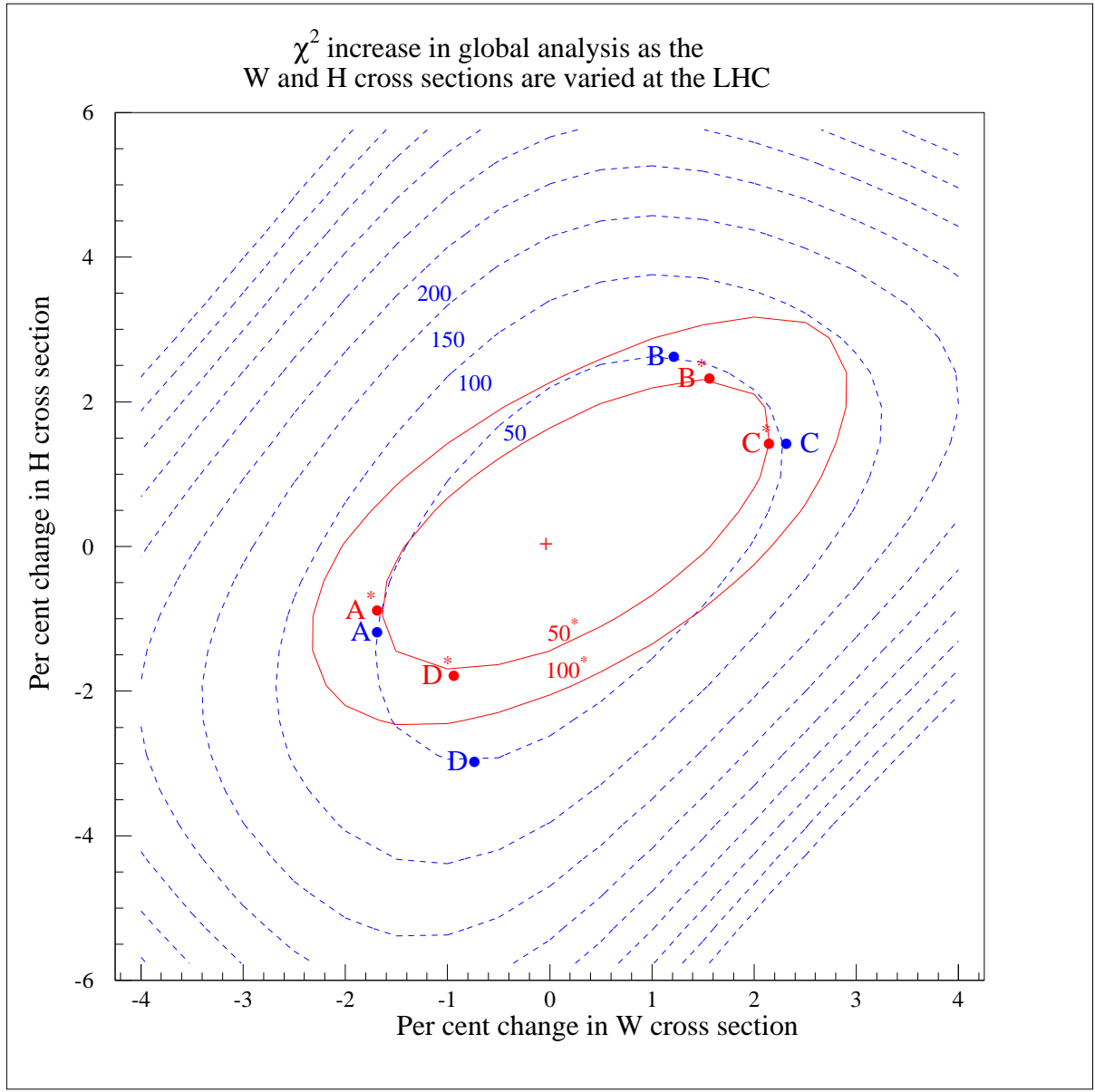,height=8in}
\vspace{-1.75cm}
 \caption{ Contours with $\Delta\chi^2=50,100\dots$ obtained by performing
global fits with the values of $\sigma_W$ and $\sigma_H$, at the LHC energy,
fixed in the neighbourhood of
their values predicted by the unconstrained MRST2001 fit. The
$\Delta\chi^2=50$ contour is taken to represent the errors on $\sigma_W$
and $\sigma_H$ (arising from the experimental errors on the data used in the
global fit). The extrema sets of partons (A,B$\dots$) are discussed in the
text. The dashed contours are obtained if $\alpha_S(M_Z^2)$ is
allowed to vary. The superimposed solid $\Delta\chi^2=50,100$ contours
are obtained if $\alpha_S(M_Z^2)$ is fixed at 0.119.}
\label{Fig7}
\end{figure}

\newpage

\begin{figure}[htbp]
\hspace{-2.5cm}
\includegraphics[width=9.5cm]{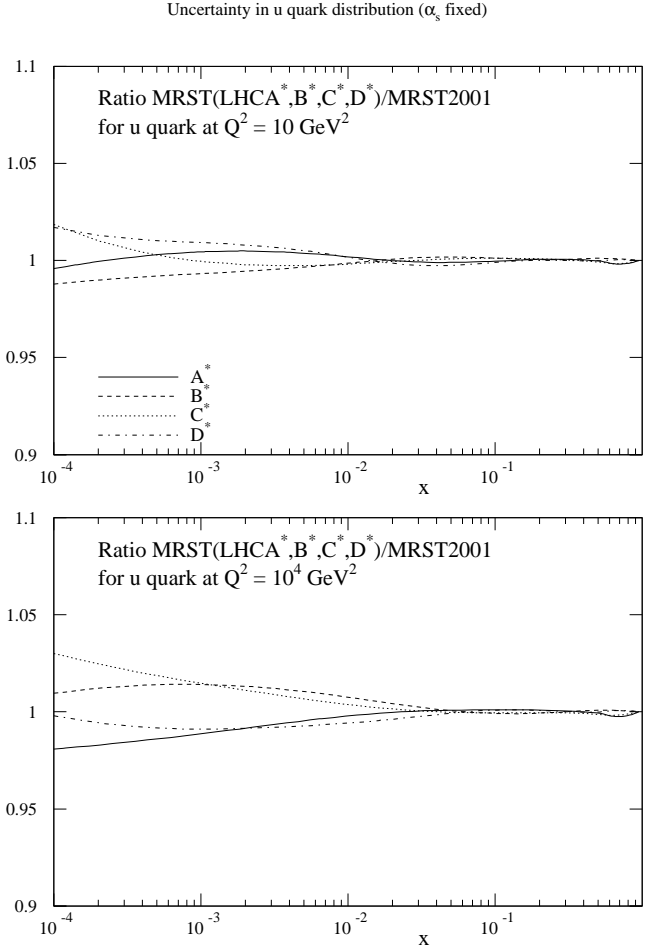}
\hspace{0.0cm}
\includegraphics[width=9.5cm]{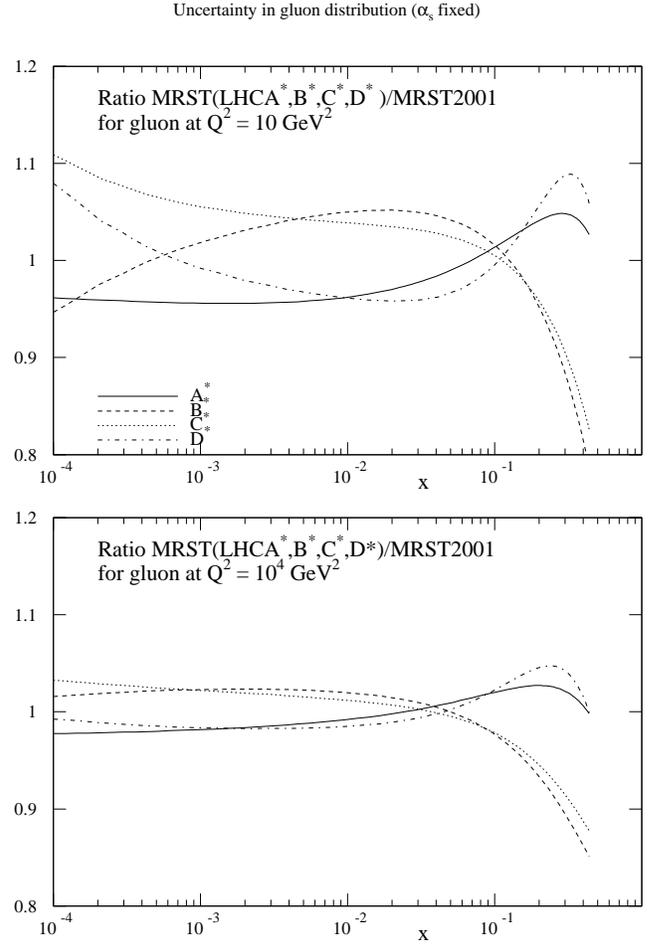}
\caption{The up quark and gluon distributions at $Q^2=10$ and
$10^4\ \GeV^2$ in the extrema global fits on the $\Delta\chi^2=50$
contour of the $\sigma_{W,H}$(LHC) plot of Fig.~7 for
$\alpha_S(M_Z^2)$ fixed at 0.119.} \label{Fig8}
\end{figure}

\newpage

\begin{figure}[htbp]
\epsfig{figure=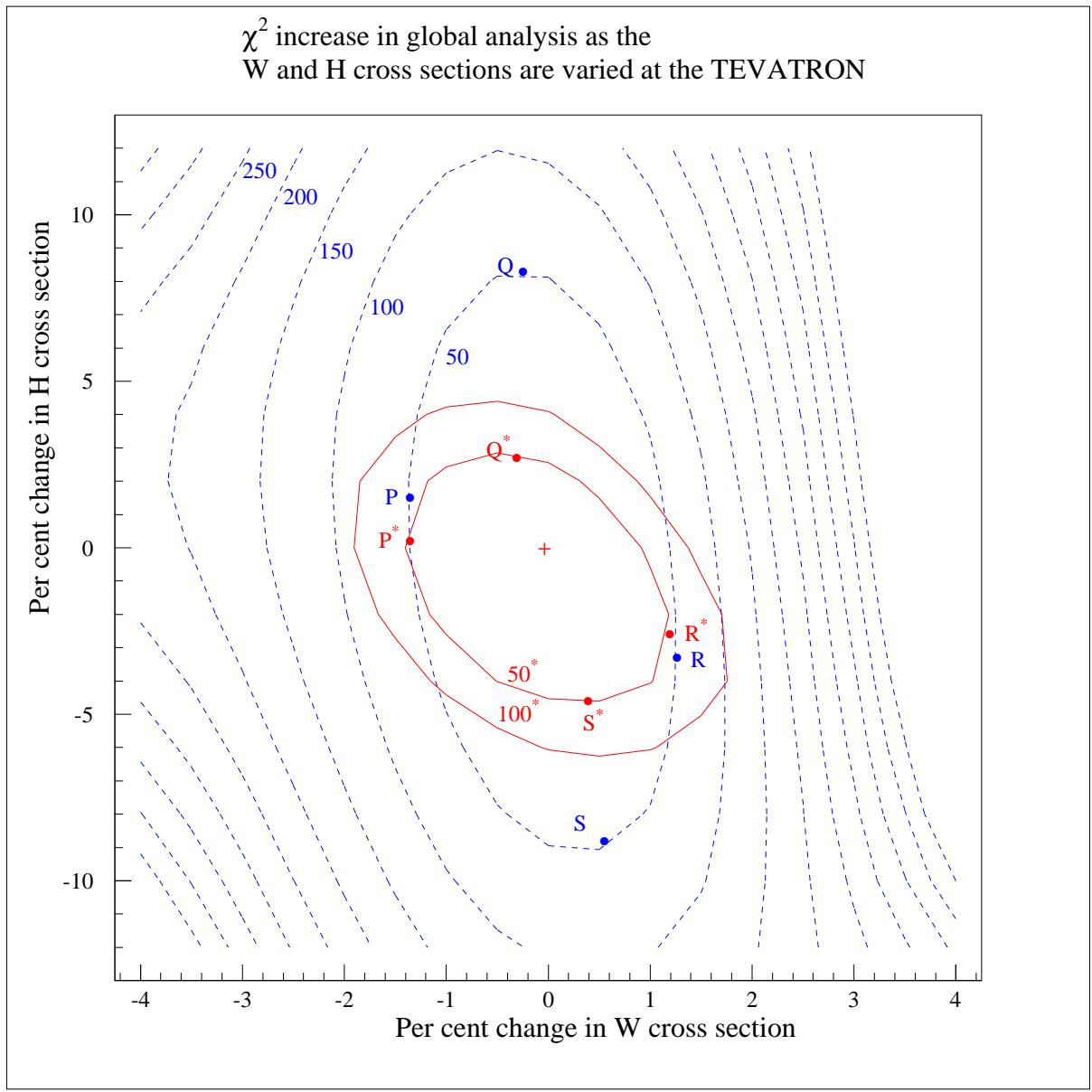,height=8in}
 \caption{As for Fig.~7, but for the Tevatron energy of $\sqrt{s}=1.8$~TeV.}
\label{Fig9}
\end{figure}

\newpage

\begin{figure}[htbp]
\hspace{-2.5cm}
\includegraphics[width=9.5cm]{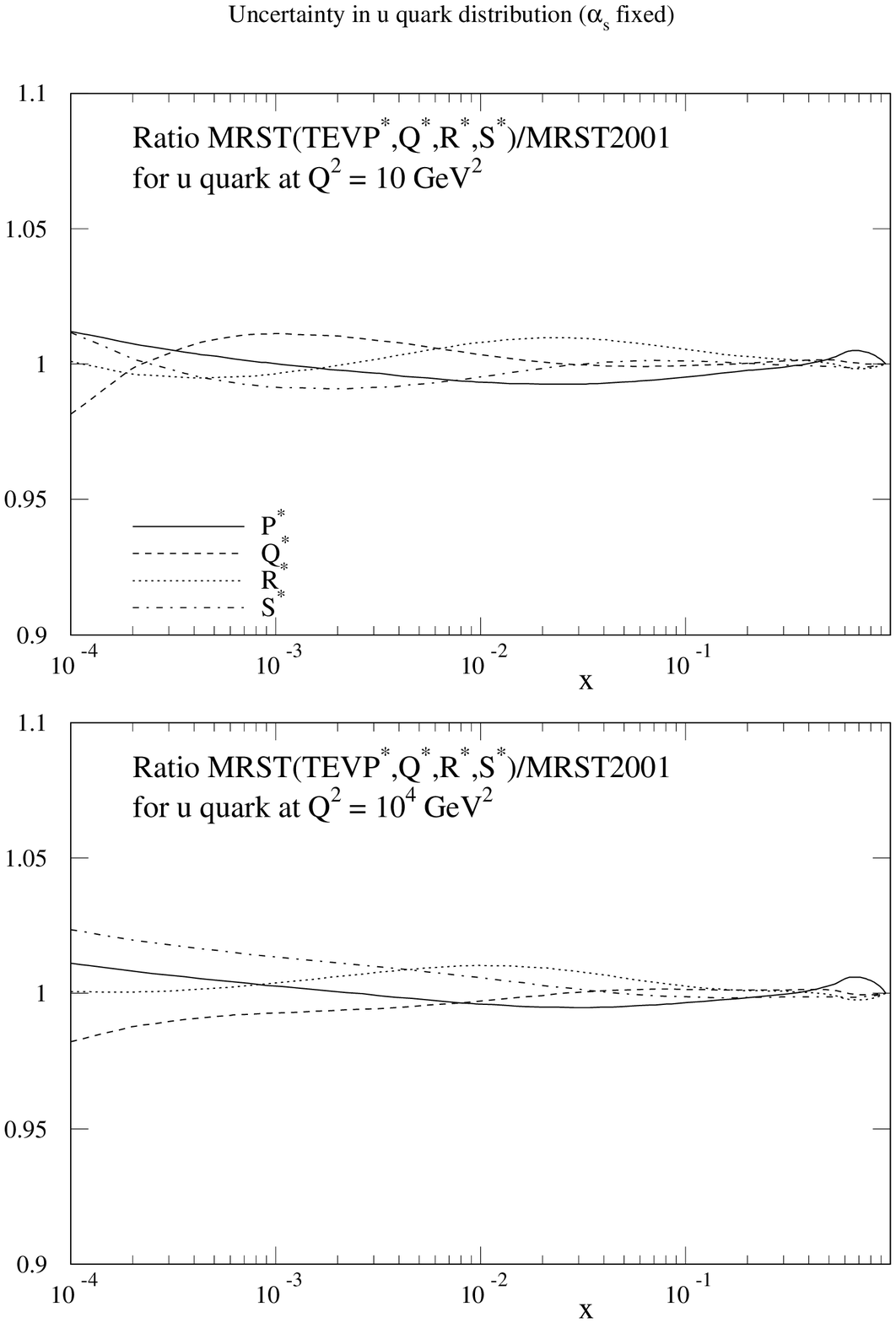}
\hspace{0.0cm}
\includegraphics[width=9.5cm]{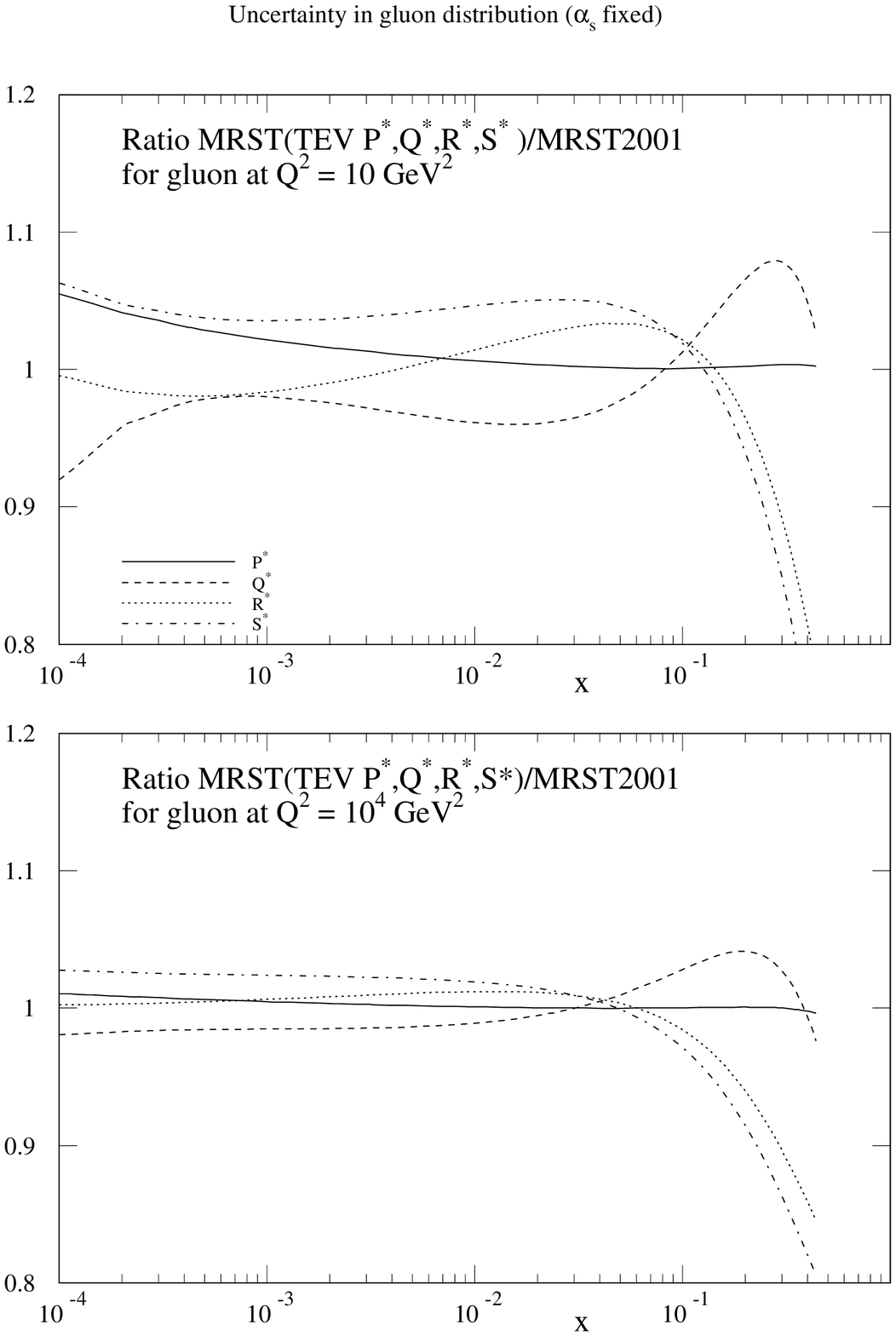}
\caption{The up quark and gluon distributions at $Q^2=10$ and
$10^4\ \GeV^2$ found in the extrema global fits on the
$\Delta\chi^2=50$ contour of the $\sigma_{W,H}$(Tevatron) plot of
Fig.~9 with $\alpha_S(M_Z^2)$ fixed at 0.119.} \label{Fig10}
\end{figure}

\newpage

\begin{figure}[htbp]
\hspace{-2.5cm}
\includegraphics[width=9.5cm]{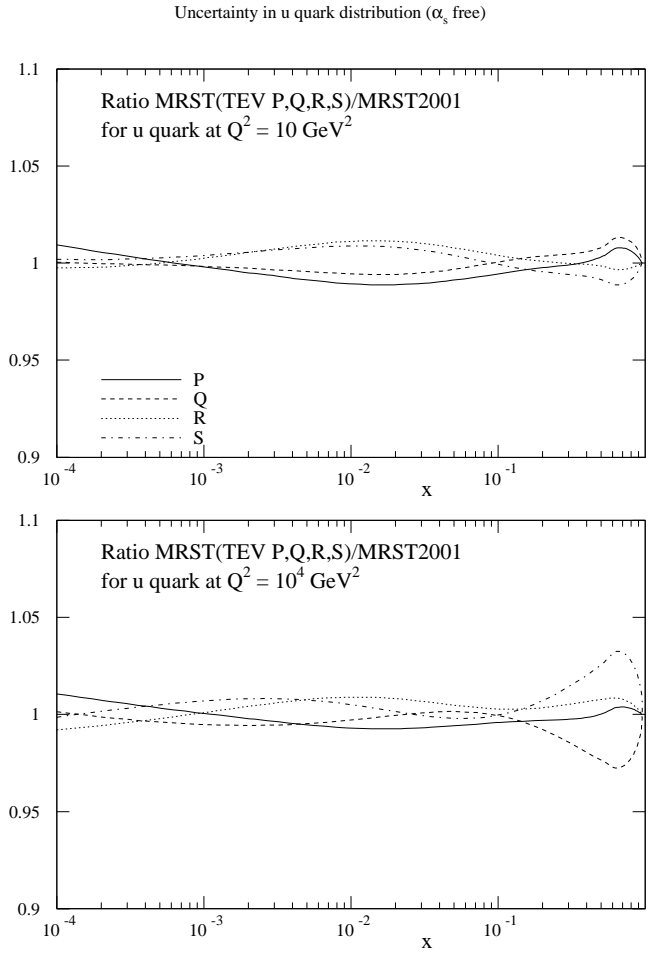}
\hspace{0.0cm}
\includegraphics[width=9.5cm]{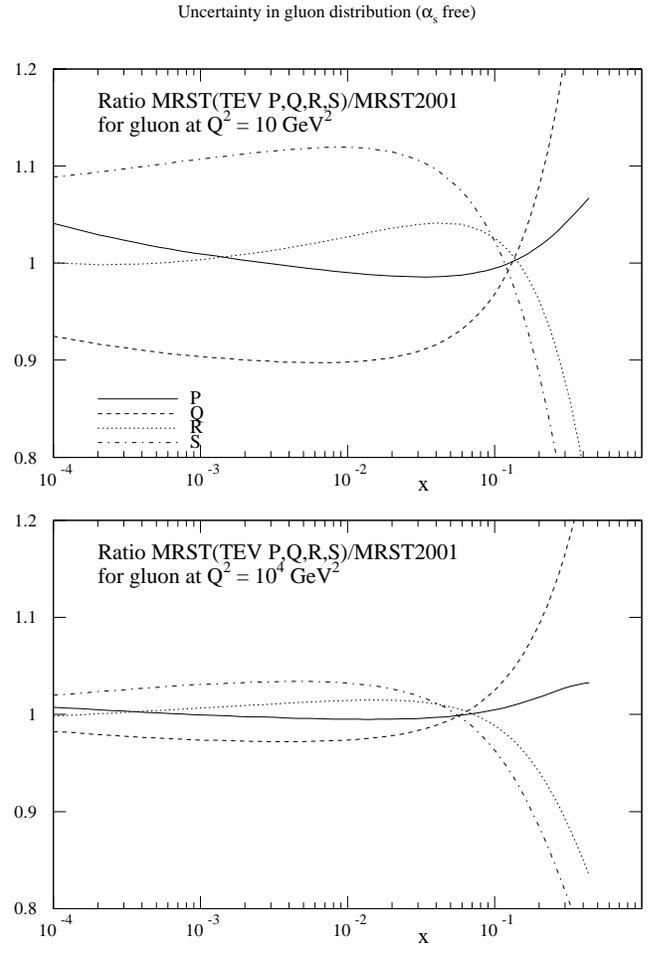}
\caption{As for Fig.~10 but with $\alpha_S(M_Z^2)$ allowed to
vary.}
\label{Fig11}
\end{figure}

\newpage

\begin{figure}[htbp]
\epsfig{figure=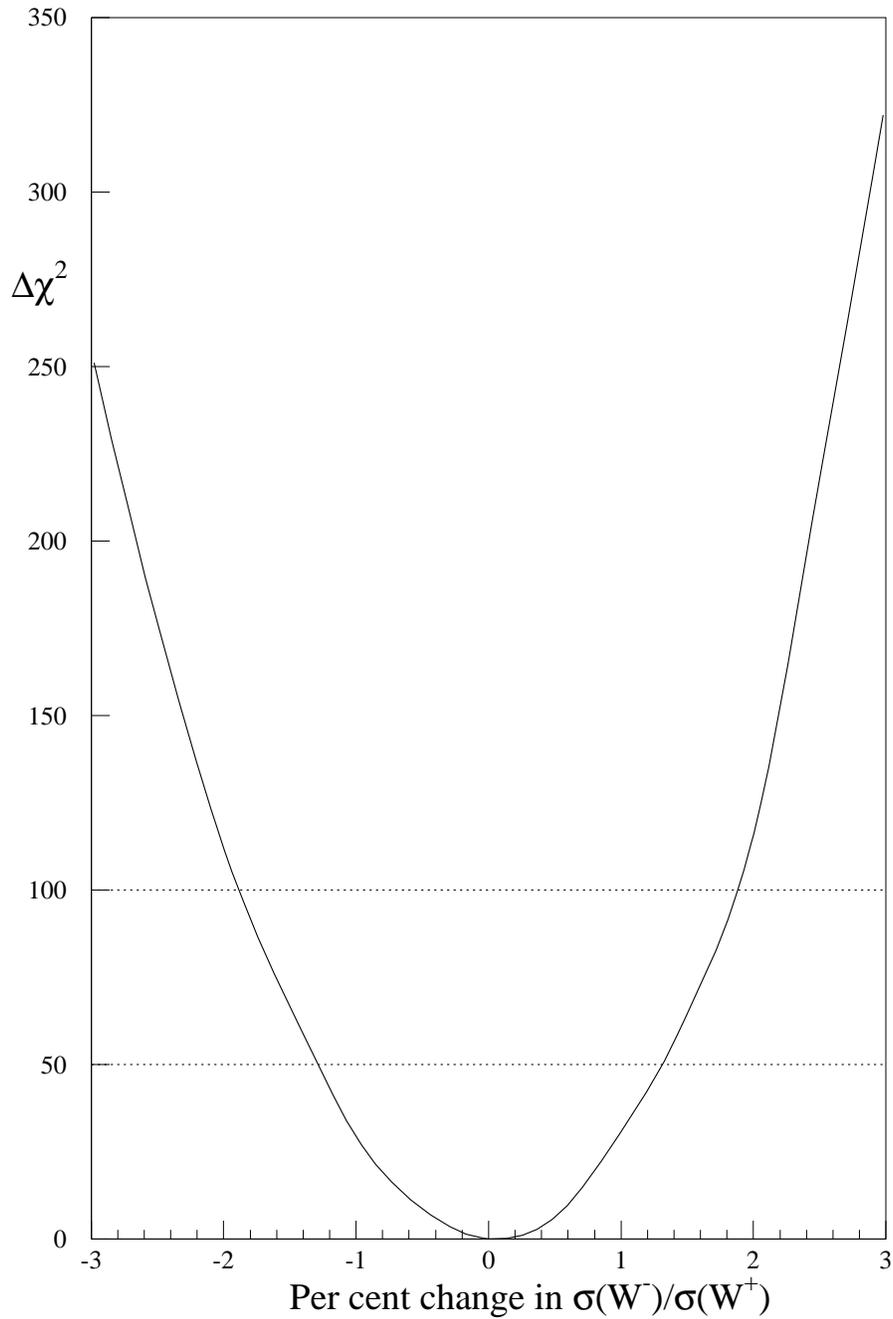,height=8in}
 \caption{The variation of $\chi^2$ obtained by performing global fits with
$\sigma(W^-)/\sigma(W^+)$ fixed at different values in the
neighbourhood of the value obtained in the unconstrained MRST2001
fit. For $\Delta\chi^2=50$ we see that the uncertainty in the
ratio is $\pm1.3\%$.} \label{Fig12}
\end{figure}

\newpage

\begin{figure}[htbp]
\epsfig{figure=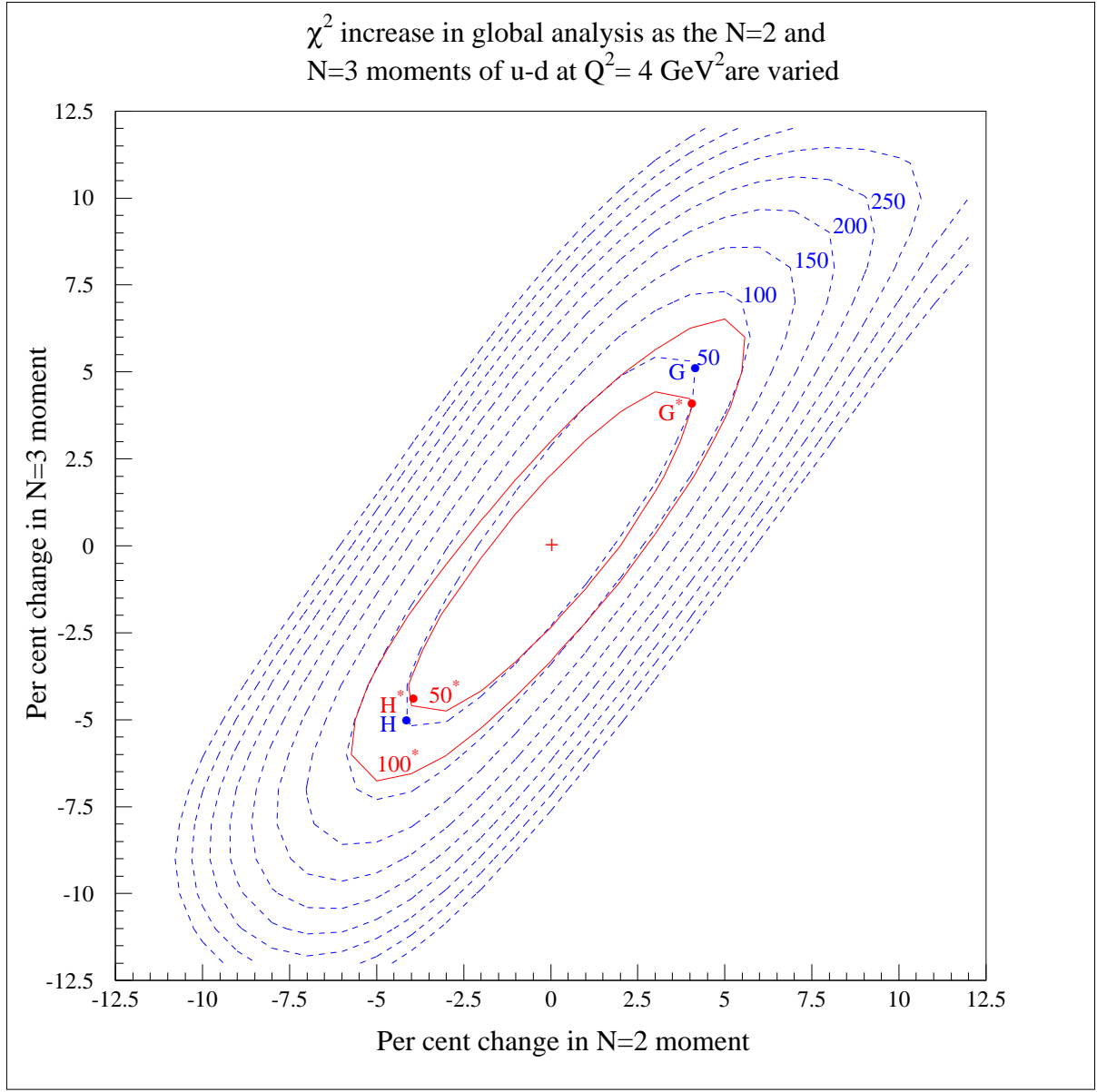,height=8in}
 \caption{The $\Delta\chi^2$ contours obtained by performing global fits with
the values of the $N=2$ and $N=3$ moments of the $u$--$d$ distribution fixed
in the neighbourhood of their values predicted by the MRST2001 global fit.
The dashed and solid curves correspond to fits with $\alpha_S(M_Z^2)$
varying and fixed respectively.}
\label{Fig13}
\end{figure}

\newpage

\begin{figure}[htbp]
\epsfig{figure=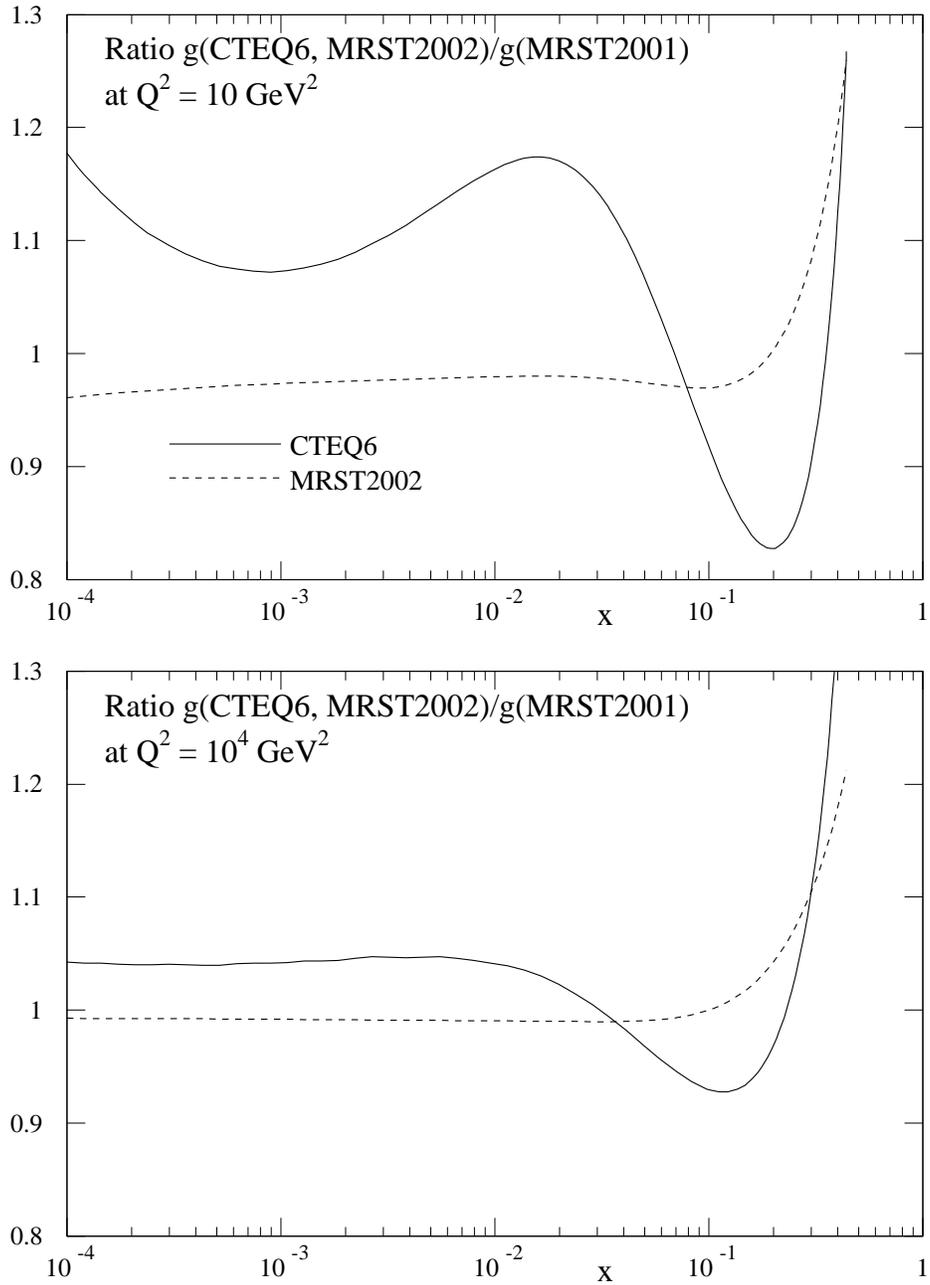,height=8in}
 \caption{The CTEQ6~\cite{CTEQ6} and MRST2002 gluon compared to
MRST2001~\cite{MRST2001} gluon at $Q^2=10$ and $10^4\ \GeV^2$.}
\label{Fig14}
\end{figure}

\newpage

\begin{figure}[htbp]
\hspace{-1.5cm}
\includegraphics[width=8cm]{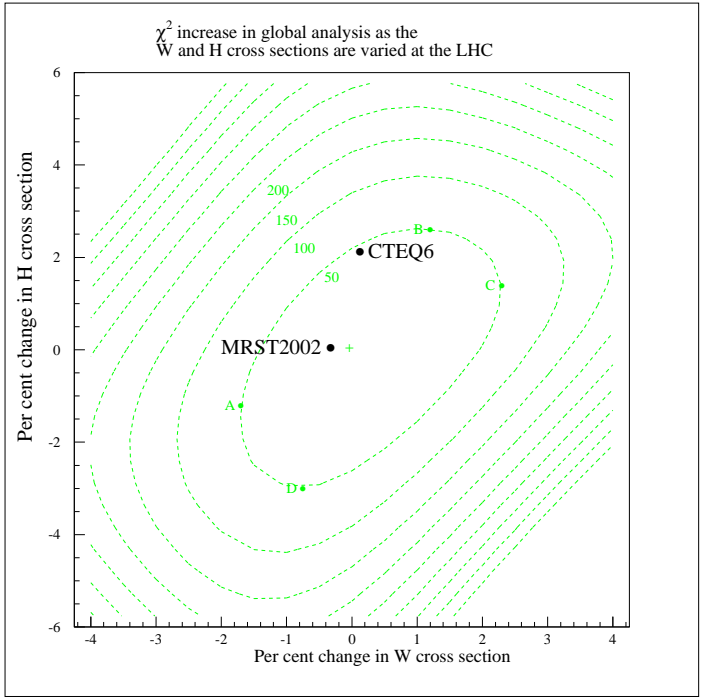}
\hspace{1.5cm}
\includegraphics[width=8cm]{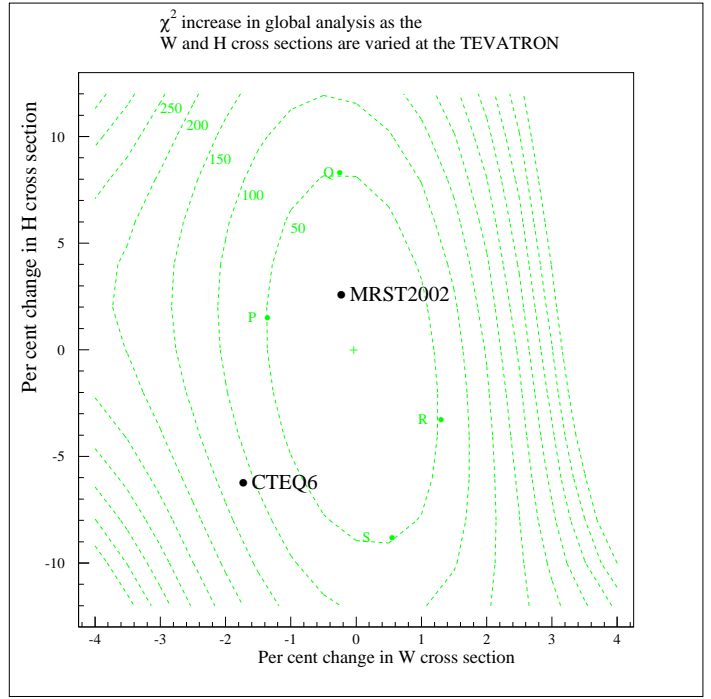}
\caption{The CTEQ6~\cite{CTEQ6} and MRST2002 predictions of
$\sigma_W$, $\sigma_H$ at the LHC and Tevatron energies, shown on
the $\Delta \chi^2$ contour plots centered on the MRST2001
partons~\cite{MRST2001}. The $\Delta\chi^2$ contours are taken
from Figs.~7 and 9 respectively, for the case in which
$\alpha_S(M_Z^2)$ is a free parameter. The inner contour with
$\Delta\chi^2=50$ is taken to represent the error on the
observables $\sigma_W$ and $\sigma_H$ arising from the {\em
experimental} errors of the data that are used in the global fit.}
\label{Fig15}
\end{figure}

\end{document}